\documentclass[aps,prd,superscriptaddress,nofootinbib]{revtex4}   
\usepackage{amssymb}
\usepackage{amsmath}
\usepackage{amssymb}
\usepackage{graphicx}
\usepackage{subfigure}
\usepackage{color}
\usepackage{mathrsfs}
\usepackage{ulem}
\usepackage[dvipsnames]{xcolor}

\usepackage[breaklinks=true,colorlinks=true]{hyperref}
\hypersetup{colorlinks=true,citecolor=blue,linkcolor=blue,urlcolor=blue}

\usepackage[utf8]{inputenc}
\usepackage[english]{babel}

\newcommand{\phiO}{\overline{\phi}}

\begin{document}

\title{Phonons scattering off discrete asymmetric solitons in the absence of a Peierls-Nabarro potential}

\author{Danial Saadatmand}
\affiliation{Institute for Theoretical Physics, Physics Department, Stellenbosch University, Matieland 7602, South Africa}
\affiliation{National Institute for Theoretical and Computational Sciences (NITheCS) South Africa}

\author{A. Moradi Marjaneh\footnote{Corresponding author}}
\affiliation{Department of Physics, Quchan Branch, Islamic Azad University, Quchan, Iran}

\author{Alidad Askari}
\affiliation{ Department of Physics, Faculty of Science, University of Hormozgan, P.O.Box 3995, Bandar Abbas, Iran}

\author{Herbert Weigel}
\affiliation{Institute for Theoretical Physics, Physics Department, Stellenbosch University, Matieland 7602, South Africa}

\begin{abstract}
We analyze the interaction of lattice vibrations (phonon wave-packets) with an asymmetric kink soliton initially at rest. 
We employ the $\phi^6$ model in one space and one time dimensions for various lattice spacings and consider two different
discretization prescriptions for the field potential that do not generate Peierls-Nabarro potentials, {\it i.e.\@} the 
kink can be placed anywhere along the lattice beyond discrete translational invariance. Since the $\phi^6$ model kink is 
neither symmetric nor anti-symmetric under spatial reflections, we simulate the cases where the wave-packet approaches the 
kink from negative or positive spatial infinity. We extract the energy transmission and reflection coefficients as functions 
of the central frequency of the phonon wave-packet for the different lattice spacings. For large lattice spacings, the wave-packet 
is always fully reflected, while for smaller spacings the amount of reflection and transmission depends on the central frequency. 
We also identify scenarios in which the target kink acquires a non-zero velocity from its interaction with the wave-packet.
\end{abstract}



\maketitle

\section{Introduction}\label{sec:introduction}
In one time and one space dimension ($D=1+1$), non-linear scalar field theories typically exhibit kink-type solutions, 
both in continuum and discrete forms. They are topologically stable because they connect different (degenerate) vacuum 
configurations at negative and positive spatial infinity. For this reason, these solutions are commonly called topological 
solitons\footnote{Strictly speaking, most of these solutions are not solitons but solitary waves. Nevertheless, it has 
become customary to use the expression {\it soliton} for localized solutions in the $\phi^6$ model \cite{Lohe:1979mh}.
Similarly, the expression {\it kink} for the $\phi^6$ model is adopted from the $\phi^4$ model \cite{Dorey:2011yw}.}.  
Soliton models in $D=1+1$ may either serve as role models for higher dimensional systems or be directly embedded
therein. In turn, they have numerous applications on all scales, ranging from cosmic strings~\cite{Nambu:1977ag} 
via hadron~\cite{Weigel:2008zz}, nuclear~\cite{Feist:2012ps} and condensed matter 
physics~\cite{Schollwock:2004aa,Nagasoa:2013} even to cosmology~\cite{Vilenkin:2000jqa}. 
The interested reader may want to consult the introduction of Ref.~\cite{MoradiMarjaneh:2022vov} for an exhaustive 
list of many specific applications.  In the context of condensed matter physics, it is particularly remarkable that 
kink-type solutions in field theory resemble the behavior of lattice kink structures in molecular dynamics 
simulations \cite{PhysRevB.103.224312}.

The interactions of $D=1+1$ solitons with each other, with impurities, with wave-packets and with small-amplitude fluctuations 
have been widely discussed~\cite{Javidan.PRE.2008,Abdelhady.IJMPA.2011,PhysRevE.94.032216,Azadeh.JHEP.2022,Alonso.PhysD.2023}. 
Recent reviews on the continuum case are, for example, Refs.~\cite{Belova:1997bq,Kevrekidis:2019} while Ref.~\cite{Braun2010-ux} 
exhaustively explores a discretized model. Discrete or discretized models are those in which the soliton is a function of 
a set of lattice points rather than a continuous variable. In $D=1+1$, that set forms a linear chain, making 
these models highly relevant to condensed matter physics \cite{BISHOP19801}. Applications based on the $\phi^4$ or 
sine-Gordon solitons are discussed in the literature \cite{Braun2010-ux,PhysRevB.76.054107}.

The lattice exhibits only discrete translation invariance, and one thus expects this to be reflected in the kink solution, 
with its position (center) constrained to only take discrete values. The force that fixes this position 
originates from the so-called Peierls-Nabarro potential (PNP). Since the continuum formulation of the kink model
does not, by construction, possess any such constraint, one is inclined to assume that discretization prescriptions
exist which lead to lattice models without such a constraint as well. The corresponding models are called PNP free.
Various such models have, for example, been explored in
Refs.~\cite{Speight:1998uq,Barashenkov_PRE,Dmitriev:2006qm,Rakhmatullina2018-kz}. Once such a prescription is adopted,
continuous translational invariance is verified by finding kink solutions for any value for the field (between the
possible vacuum values) at an arbitrary lattice point.

In most cases the continuum models have a Bogomolny-Prasad-Sommerfield (BPS) \cite{Bogomolny:1975de,Prasad:1975kr}
construction for the static soliton $\phi(x)$: Let $V(\phi)$ be the field potential, then the soliton energy is 
minimized by solutions to the first order equation $\frac{\partial \phi(x)}{\partial x}=\pm\sqrt{2V(\phi)}$ where 
the two signs refer to kink and antikink constructions. Another interesting feature of the BPS construction is 
that the (classical) energy is fully determined by the boundary values of the field. When applied to impurity 
models~\cite{Adam:2019djg}, this results in degenerate solutions that differ in shape. That degeneracy
is removed only on the quantum level \cite{Evslin:2022xmp,Takyi:2022sng,Graham:2022rqk}.

The discretized analog of this first order equation is $\phi_n-\phi_{n-1}=hu_n$, where $\phi_n$ is the field value 
at lattice side $n$, and $h$ is the lattice spacing. The right hand side is a function of the field values in 
the vicinity of $n$ so that $u_n\,\to\,\pm\sqrt{2V(\phi)}$ when $h\,\to\,0$. A construction principle for a PNP free 
discretization is to write $u_n$ in terms of $\phi_n$ and $\phi_{n-1}$ such that it is invariant under the exchange 
of these two function values \cite{Dmitriev:2006qm}. In the current study, we will consider two distinct discretization
prescriptions for the $\phi^6$ model, which has the field potential $V(\phi)=\frac{k}{8}\phi^2\left(1-\phi^2\right)^2$,
where $k$ is a positive coupling constant. One prescription is taken from Ref.~\cite{Rakhmatullina2018-kz}, while the 
other is new. In contrast to the prescription from Ref.~\cite{Rakhmatullina2018-kz} the novel one is a realization of 
the BPS construction in the discretized $\phi^6$ model. 

Our main goal is to determine the translationally invariant kink and then consider the scattering of phonons off it.
Phonons are considered as wave-packets built from small amplitude fluctuations about the kink. The $\phi^6$ model is 
particularly interesting because the kink, or more precisely, its energy density, is not invariant under spatial 
reflection. Hence, we expect different scattering data for wave-packets approaching the kink from the left or right. 
Such asymmetries have recently been observed for kinks interacting with an impurity that models a resistance 
force~\cite{Moradi.EPJB.2022}.  In the $\phi^6$ model, the kink connects $\phi=0$ and $|\phi|=1$ between the two 
spatial infinities. In the continuum version of the model, the classical phonon back-reaction causes the kink to 
move towards the regime with $|\phi|=1$~\cite{Romanczukiewicz:2017hdu}. In the discrete version, we thus not only 
need to analyze the difference of scattering data for wave-packets approaching the kink from the left or right but 
we also explore whether the motion of the kink itself is as distinct as in the continuum case. Due to the 
asymmetry, the dispersion relations for the small amplitude fluctuations are different for the two opposite 
sides of the kink. Thus, we can interpret the discrete $\phi^6$ soliton as modeling a smooth transition between 
similar but nonetheless different crystal lattices. As an intermediate result, we find energy bands for the small 
amplitude fluctuations about the $\phi^6$ model kink. Such spectral results are of general interest, for example, 
they are used to construct the phenomenological free energy~\cite{PhysRevB.22.477} in statistical mechanics or to 
compute the vacuum polarization energy of localized configurations~\cite{Graham:2009zz} in quantum field theory.

Let us briefly mention some of the progress made on lattice kinks in addition to what has been reported in the above 
quoted papers. Kink-antikink collisions have been explored in the discrete $\phi^4$ model \cite{ASKARI2020109854}. An 
anhormonically driven chain (without kink) was considered in Ref.~\cite{PhysRevE.97.022217}. That study also explored 
the role of the breather, as did Ref.~\cite{Evazzade2018-fi}. The authors of Ref.~\cite{PhysRevE.52.R2183} investigated
the propagation (with damping) of a kink along a chain. Solitons in PNP free $\phi^4$ models were, to our knowledge, first 
found by Speight \cite{Speight1997-hc}. There have been earlier studies of the phonon kink interaction. For example, the 
interaction of large amplitude (anharmonic) phonons with the sine-Gordon kink was investigated in Ref.~\cite{Jaworski.PLA.1987}. 
Another example is the interaction of phonons with the standing discrete breather solitons which was discussed for hard- 
and soft-type anharmonicities in Ref.~\cite{Hadipour.PLA.2020}.

This paper is organized as follows: In Sec.~\ref{sec:phi6model} we introduce the model and present the theory needed 
to discretize the $\phi^6$ model. This will be followed by the presentation and discussion of the numerical results 
in  Sec.~\ref{Phonon-kink}. We summarize with concluding remarks in Sec.~\ref{sec:conclusions}. Some technicalities 
are relegated to two appendices.

\section{The model}\label{sec:phi6model}
In this section, we present our model in $D=1+1$ dimensions. We start by reviewing the continuum version and subsequently
introduce its discretized counterparts. 

The general form of the continuum Lagrangian for the scalar field $\phi(x,t)$ with a sixth order field potential that 
allows for spatially asymmetric static solutions is\footnote{The more general form with a field potential proportional 
to $(\phi^2+a^2)(\phi^2-v^2)^2$ only has symmetric solutions \cite{Lohe:1979mh}.}
\begin{equation} \label{eq:lagrang1}
\mathcal{L}=\frac{1}{2}\left(\partial_t\phi\right)^2 -\frac{1}{2}\left(\partial_x\phi\right)^2-
\frac{k}{8}\phi^2(\phi^2-v^2)^2\,.
\end{equation}
Here $k>0$ is a coupling constant and $v$ is a (possible) vacuum expectation value of $\phi$. Scaling the field by
$\phi\,\to\,v\phi$ and introducing dimensionless coordinates according to
$(t,x)\,\to\,(t,x)\left[\frac{1}{v^2}\sqrt{\frac{2}{k}}\right]$
moves the dependence on the model parameters into an overall factor
\begin{equation} \label{eq:lagrangy}
\mathcal{L}=\frac{kv^6}{2}\left[\frac{1}{2}\dot{\phi}^2-\frac{1}{2}\phi^{\prime2}-\frac{1}{4}\phi^2(\phi^2-1)^2\right]\,,
\end{equation}
where dots and primes denote derivatives with respect to the new, dimensionless time and space variables, respectively. 
We will adopt that notation henceforth. The Euler Lagrange equation 
\begin{equation}
\ddot{\phi}-\phi^{\prime\prime}+\frac{1}{2}\phi(1-4\phi^2+3\phi^4)=0\,.
\label{eq:Phi6EOM}
\end{equation}
associated with these variables does not contain the model parameters and we can make parameter independent and
universal statements.

This field equation has well-known static solutions\footnote{For simplicity we use identical symbols
for the static and time-dependent solutions.}
\begin{equation}
\phi_{(0,\pm1)}(x)=\pm \sqrt{\frac{1}{2}\left[1+\tanh\frac{x}{\sqrt{2}}\right]}
\quad{\rm and}\quad
\phi_{(\pm1,0)}(x)=\pm \sqrt{\frac{1}{2}\left[1-\tanh\frac{x}{\sqrt{2}}\right]}\,,
\label{eq:Phi6Solution1}
\end{equation}
that fall into topological sectors characterized by the field values at negative and positive spatial infinity and
there are four distinct solutions that cannot continuously be deformed into one another. 
More general and time-dependent solutions to Eq.~(\ref{eq:Phi6EOM}) are constructed by a combination 
of a translation and a Lorentz-boost:~$\phi(x)\,\mapsto \phi\left(\frac{x-x_0-vt}{\sqrt{1-v^2}}\right)$.
More interestingly, we note that the classical energy functional
\begin{equation}
E_{\rm cl}[\phi]=\frac{1}{2}\int_{-\infty}^{+\infty}dx\,
\left[\phi^{\prime2}+\frac{1}{2}\phi^2(1-\phi^2)^2\right]
=\frac{1}{2}\int_{-\infty}^{+\infty}dx\,\left[\phi^\prime\pm\frac{1}{\sqrt{2}}\phi(1-\phi^2)\right]^2
\mp\frac{1}{4\sqrt{2}}\phi^2\left[2-\phi^2\right]_{-\infty}^{\infty}\
\label{eq:BPS1}
\end{equation}
has a BPS construction with first order field equations $\pm\phi^\prime-\frac{1}{\sqrt{2}}\phi(1-\phi^2)=0$.
The four solutions in Eq.~(\ref{eq:Phi6Solution1}) are transformed into one another by swapping the overall 
sign and/or spatial reflection. It is then sufficient to consider only one of the two BPS constructions, 
specifically we define $u(\phi,\phi^\prime)=\phi^\prime-\frac{1}{\sqrt{2}}\phi(1-\phi^2)=0$, which leads to 
$\phi_{(0,\pm1)}(x)$. Numerically that first order differential equation is solved by picking an arbitrary 
position $x_0$ and prescribing $\phi(x_0)\in[-1,1]$. Selecting $\phi(x_0)\in[0,1]$ produces $\phi_{(0,1)}(x)$ 
while $\phi(x_0)\in[-1,0]$ leads to $\phi_{(0,-1)}(x)$. The profiles $\phi_{(\pm1,0)}(x)$ originate from the 
alternative BPS construction. Choosing one of the two BPS constructions and fixing the sign of the field 
at a particular point uniquely defines the static solution.

To proceed to the discretized version, we note that, up to surface contributions, the Hamilton functional 
is\footnote{With this sign convention the integral is a lower bound for configurations with $\phi^\prime\ge0$.}
\begin{equation}
\mathcal{H}=\frac{1}{2}\int dx\left[\pi^2+u^2(\phi,\phi^\prime)\right]
\quad{\rm where}\quad \pi=\dot{\phi}
\quad{\rm and}\quad u(\phi,\phi^\prime)=\phi^\prime-\frac{1}{\sqrt{2}}\phi(1-\phi^2)\,.
\label{eq:Hamiltonian}
\end{equation}

We next turn to discretized versions by introducing an equi-distant lattice $x_n=nh$ where 
$n=-N,-N+1,\ldots,-1,0,1,\ldots,N$. We stress that this spacing refers to the dimensionless
coordinate. Returning to the physical coordinate requires devision by $v^2\sqrt{k/2}$.
Let us first consider \cite{Rakhmatullina2018-kz}
\begin{equation}
\phi_n=\phi(x_n,t)\,,\quad
u_n=\frac{\phi_n-\phi_{n-1}}{h}-\frac{1}{\sqrt{2}}\frac{\phi_{n-1}+\phi_n}{2}
\left(1-\frac{\phi^2_{n-1}+\phi_{n-1}\phi_n+\phi^2_n}{3}\right)
\quad{\rm and}\quad
\dot{\phi}_n=\dot{\phi}(x_n,t).
\label{eq:discrete1}
\end{equation}
In the continuum limit, the first term in $u_n$ turns into $\phi^\prime$, while the second term approaches
$-\frac{1}{\sqrt{2}}\phi(1-\phi^2)$. According to Refs.~\cite{Speight1997-hc,Speight:1998uq}
we write the discretized Hamiltonian as
\begin{equation}
\mathcal{H}=\frac{h}{2} \sum\limits_{n=1-N}^N\left(\pi_n^2+u_n^2 \right)\,.
\label{eq:discretHamiltonian}
\end{equation}
The static equation is $u_n=0$ for all $n$. To generate a kink configuration, we prescribe $\phi_n$
and compute $\phi_{n-1}$ from that equation. Similarly, we get $\phi_{n+1}$ from $u_{n+1}=0$ when $\phi_n$ 
is given. We combine these two equations in writing
\begin{equation}
\frac{1}{12}\phi_{n\pm1}^3+\frac{1}{6}\phi_n\phi_{n\pm1}^2
+\left(\pm\frac{1}{\sqrt{2}h}-\frac{1}{4}+\frac{1}{6}\phi_n^2\right)\phi_{n\pm1}
+\left(\mp\frac{1}{\sqrt{2}h}-\frac{1}{4}+\frac{1}{12}\phi_n^2\right)\phi_n=0\,.
\label{eq:Phi6kinksolution}
\end{equation}
Each of these cubic equations typically has one real solution for each $\phi_{n-1}$ and $\phi_{n+1}$ when 
$\phi_n\in[0,1]$ is prescribed. In the next step, we 
extract $\phi_{n-2}$ and $\phi_{n+2}$ from $u_{n-1}=0$ and $u_{n+2}=0$, respectively. In repeating this 
procedure we cover the whole lattice. According to the discussion above, this will produce a kink configuration 
with $\phi_{-N}\approx0$ and  $\phi_N\approx1$ provided the initial value is far enough away from the boundaries.
We refer to this discretized kink solution as $\phi_n^{(0)}$. Typically, we place the kink in the center, {\it i.e.\@}
we initiate the sequence with $\phi_{0}\in[0,1]$. The larger we take $\phi_0$, the further to the left the kink 
is localized. The fact that we can assume any value in this interval reflects translational invariance, and 
there is no potential barrier for displacing the localized soliton. The absence of the barrier is usually framed 
as a scenario being free of a Peierls-Nabarro potential.

We also want to explore the relation to the BPS construction. To this end, we need 
\begin{align}
u_n^2&=\left(\frac{\phi_n-\phi_{n-1}}{h}\right)^2
+\frac{1}{2}\left(\frac{\phi_{n-1}+\phi_n}{2}\right)^2
\left(1-\frac{\phi^2_{n-1}+\phi_{n-1}\phi_n+\phi^2_n}{3}\right)^2
\cr  & \hspace{1cm}
\label{eq:DBPS1}
-\frac{1}{\sqrt{2}h}\left(\phi_n^2-\phi_{n-1}^2\right)
\left(1-\frac{\phi^2_{n-1}+\phi_{n-1}\phi_n+\phi^2_n}{3}\right)\,.
\end{align}
The first two terms are associated with $\phi^{\prime2}$ and the field potential $V(\phi)$, respectively. 
A BPS constructions entails that, when summing over $n$ the contribution from last term only contains
the field at the boundaries, {\it i.e.\@} $\phi_{-N}$ and $\phi_N$. However, the explicit calculation yields
\begin{equation}
\sum_{n=1-N}^N \left(\phi_n^2-\phi_{n-1}^2\right)
\left(1-\frac{\phi^2_{n-1}+\phi_{n-1}\phi_n+\phi^2_n}{3}\right)
=\phi_N^2-\phi_{-N}^2-\frac{1}{3}\left(\phi_N^4-\phi_{-N}^4\right)
-\frac{1}{3}\sum_{n=1-N}^N\left(\phi_n^3\phi_{n-1}-\phi_{n-1}^3\phi_{n}\right)\,.
\label{eq:DBPS2}
\end{equation}
Obviously, the last sum cannot be reduced to boundary contributions. Therefore, we also consider
a second, alternative discretization prescription,
\begin{equation}
\overline{u}_n=\frac{\phi_n-\phi_{n-1}}{h}
-\frac{1}{\sqrt{2}}\frac{\phi_{n-1}+\phi_n}{2}\left(1-\frac{\phi^2_{n-1}+\phi^2_n}{2}\right),
\label{eq:NEW}
\end{equation}
which by a similar calculation turns the analog of the last term in Eq.~(\ref{eq:DBPS1}) indeed into a 
boundary expression as required for the BPS construction.\footnote{For the field potential $(1-\phi^2)^2$ 
the discretization analog to Eq.~(\ref{eq:discrete1}), {\it i.e.} 
$(1-\phi^2)\,\to\,1-\frac{\phi^2_{n-1}+\phi_{n-1}\phi_n+\phi^2_n}{3}$ is BPS consistent.}
This discretization yields the cubic equation
\begin{equation}
\frac{1}{8}\,\phiO_{n\pm1}^3+\frac{1}{8}\,\phiO_n\phiO_{n\pm1}^2
+\left(\pm\frac{1}{\sqrt{2}h}-\frac{1}{4}+\frac{1}{8}\phiO_n^2\right)\phiO_{n\pm1}
+\left(\mp\frac{1}{\sqrt{2}h}-\frac{1}{4}+\frac{1}{8}\phiO_n^2\right)\phiO_n=0\,.
\label{eq:Phi6kinksolutionNEW}
\end{equation}

To explore the interaction of that kink with a wave-packet (or a dislocation), we require the full equations
of motion for the discretized version of the model. We apply the variation principle $\phi_n\,\to\,\phi_n+\delta_n$
with $\delta_1=\delta_N=0$ to the potential part of the Hamiltonian
\begin{align}
\label{eq:expandpot}
\frac{1}{2}\sum_n u_n^2\,\to\,\sum_{n} U_n[\phi] \delta_n\qquad{\rm with}\qquad
U_n[\phi]&=\sqrt{2}\left[\frac{1}{\sqrt{2}h}-\frac{1}{4}
+\frac{1}{12}\left(2\phi_{n-1}^2+4\phi_n\phi_{n-1}+3\phi_n^2\right)\right]u_n\\
\nonumber &\hspace{0.5cm}-\sqrt{2}\left[\frac{1}{\sqrt{2}h}+\frac{1}{4}
-\frac{1}{12}\left(3\phi_n^2+4\phi_n\phi_{n+1}+2\phi_{n+1}^2\right)\right]u_{n+1}\,.
\end{align}
The field equations then read
\begin{equation}
\ddot{\phi}_n(t)=-U_n[\phi]\,.
\label{eq:fieldequation}
\end{equation}
Wave-like solutions emerge from linearizing that equation in the small amplitude fluctuation
$\epsilon_n(t)$, defined as the deviation from the solution to Eq.~(\ref{eq:Phi6kinksolution}):
\begin{equation}
\phi_n(t)=\phi_n^{(0)}+\epsilon_n(t)
\qquad\mbox{so that}\qquad
\ddot{\epsilon}_n(t)=\sum_{m=-1}^1 V_n^{(m)}[\phi^{(0)}]\epsilon_{n+m}(t).
\label{eq:fluctuation}
\end{equation}
Explicit expressions for the coefficient functions $V_n^{(m)}$ in terms of the discretized kink are 
listed in Appendix \ref{appendix} for both discretization prescriptions, as is the analog of 
Eq.~(\ref{eq:fieldequation}) for $\overline{u}_n$.

We have set up the BPS procedure such that $\phi_n^{(0)}\sim0$ for small $n$ and $\phi_n^{(0)}\sim1$
when $n{\scriptstyle \lesssim}|N|$. For these cases, we get from the formulas in Appendix \ref{appendix}
\begin{equation}
V_n^{(0)}[0]=-\frac{2}{h^2}-\frac{1}{4}\,,\quad V_n^{(\pm1)}[0]=\frac{1}{h^2}-\frac{1}{8}
\quad {\rm and}\quad
V_n^{(0)}[\pm1]=-\frac{2}{h^2}-1\,,\quad V_n^{(\pm1)}[\pm1]=\frac{1}{h^2}-\frac{1}{2}
\label{eq:Vasymp}
\end{equation}
for either of the two discretization prescriptions as they have identical $V_n^{(m)}$ when
$\phi_{n-1}=\phi_n=\phi_{n+1}$. Employing the standard parameterization for one-dimensional lattice vibrations, 
$\epsilon_n(t)\propto{\rm exp}\left[{\rm i}\left(nqh-\omega t\right)\right]$ yields the (free) 
dispersion relations
\begin{equation}
\omega^2=\omega_1^2=\left(\frac{4}{h^2}-\frac{1}{2}\right)\sin^2\frac{qh}{2}+\frac{1}{2}\quad 
 \mbox{for}\quad \phi\equiv0 
\qquad {\rm and}\qquad
\omega^2=\omega_2^2=\left(\frac{4}{h^2}-2\right)\sin^2\frac{qh}{2}+2\quad 
\mbox{for}\quad |\phi|\equiv1 ,
\label{eq:Phi6Spectrom}
\end{equation}
where $\omega$ is the frequency and $q$ is the wave-number. Note that this frequency and the wave-number
are defined with respect to the dimensionless coordinates. The emergence of two distinct dispersion 
relations causes different frequency bands on either side of the kink. This is inherited from the 
different curvatures of the $\phi^6$ model potential at the degenerate minima $\phi=0$ and $\phi=\pm1$. 
In the continuum limit, this corresponds to different masses at positive and negative spatial infinity 
and leads to instabilities on the quantum level 
\cite{Weigel:2016zbs,Weigel:2017iup,Romanczukiewicz:2017hdu}\footnote{See Sect. 3
of Ref.~\cite{Graham:2022rqk} for a recent review on quantum corrections to soliton energies in $D=1+1$.}. 
However, here we are only interested in classical solutions. 

\begin{figure*}[ht!]
\begin{center}
  \centering
  \subfigure[]
{\includegraphics[width=0.3
 \textwidth]{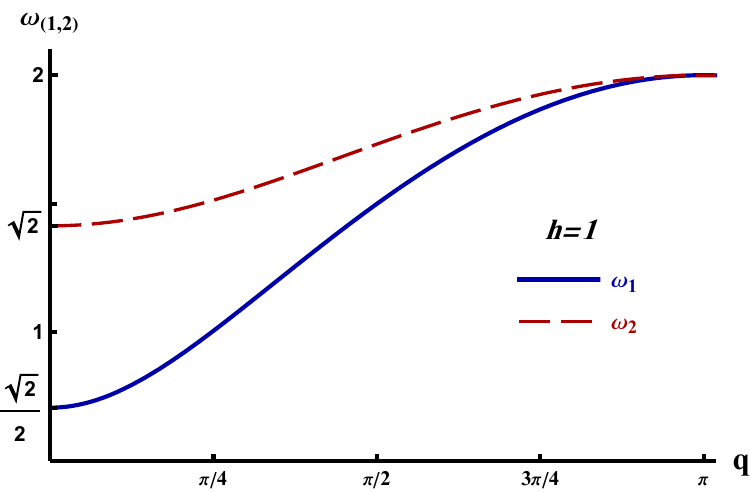}\label{fig:omega_q_h1}}
\subfigure[]{\includegraphics[width=0.3
 \textwidth]{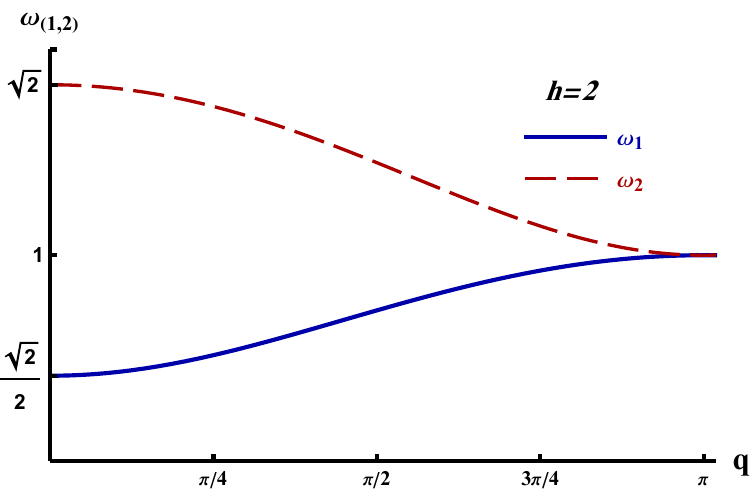}\label{fig:omega_q_h2}}
  \subfigure[]{\includegraphics[width=0.3
 \textwidth]{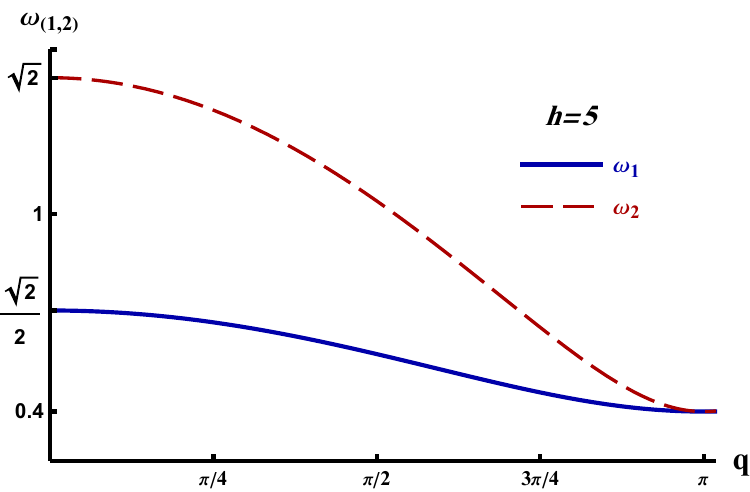}\label{fig:omega_q_h5}}
  \\
\caption{Dispersion relations for small-amplitude waves (phonons), Eq.~(\ref{eq:Phi6Spectrom}) for different 
lattice spacings $h$.}
  \label{fig:Dispersion}
\end{center}
\end{figure*}

The above dispersion relations have a number of interesting features. For small lattice spacings, $h\,\to\,0$ 
we recover the continuum relations $\omega_1^2\approx q^2+1/2$ and $\omega_2^2\approx q^2+2$ which are monotonously 
increasing functions of the wave-number. We may thus call these dispersion relations acoustic\footnote{They are 
acoustic in the sense that the frequency is an increasing function of the wave-number. However, due to the relativistic 
formulation, they are not acoustic in the sense that the frequency is a linear function at small wave-numbers.}.
As $h$ is increased, the coefficient of the sine-function eventually becomes negative and the frequency decreases 
as the wave-number increases, {\it i.e.} the dispersion relations become optic. A major reason that they turn 
optic at larger lattice spacings is the particular discretization description for the non-PNP scenario. The
prescription leads to the $h$ independent entries in the coefficients of $\sin^2(qh/2)$.
In appendix \ref{appendixB}, we corroborate on this argument by exploring a suggestive discretization prescription 
that has a PNP but produces an acoustic dispersion relation for all $h$. At the edge of the Brillouin zone, $|q|=\pi/h$, 
the frequency even approaches zero when $h\,\to\,\infty$. Also at the edge of the Brillouin zone, the two frequencies 
are equal. These features are displayed in Fig.~\ref{fig:Dispersion}.

\begin{figure*}[ht!]
\begin{center}
  \centering
      \subfigure[]{\includegraphics[width=0.3
 \textwidth]{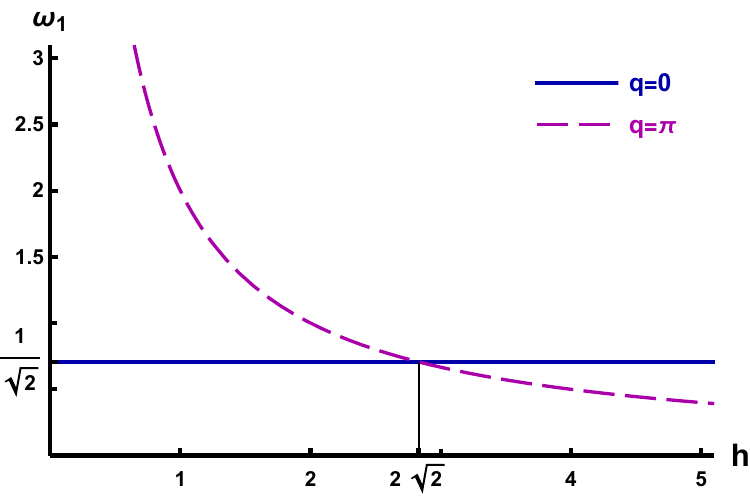}\label{fig:omega1h}}
   \hspace{2cm}
  \subfigure[]{\includegraphics[width=0.3
 \textwidth]{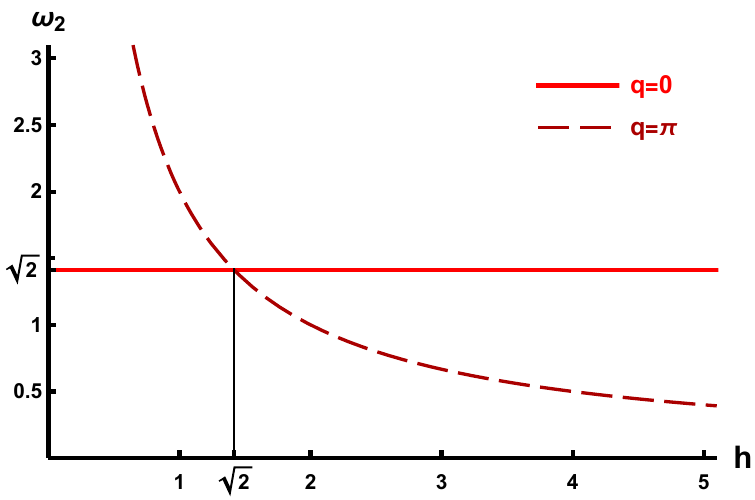}\label{fig:omega2h}}
  \\
\caption{Boundaries of the phonon spectrum as functions of $h$, Eq.~(\ref{eq:Phi6Spectrom}), 
(a) $\omega_1$ and (b) $\omega_2$.}
  \label{fig:omegah}
\end{center}
\end{figure*}

The frequencies are functions of the wave-number $q$ and the lattice spacing~$h$. Since the sine-function is 
bounded, the frequencies are so too. For any given $h$, the values $q=0$ and $q=\frac{\pi}{h}$ determine the 
frequency interval for which phonon modes do exist. For $\omega_1$, the bounds are $\frac{1}{\sqrt{2}}$ and 
$\frac{2}{h}$ while $\omega_2$ is bounded by $\sqrt{2}$ and~$\frac{2}{h}$. For small $h$ the upper bound is 
$\frac{2}{h}$ for both frequencies. As shown in Fig.~\ref{fig:omegah}, the roles of upper and lower bounds switch 
when increasing $h$ since the coefficients of the sine-functions become negative at $h=2\sqrt{2}$ for $\omega_1$ 
and at $h=\sqrt{2}$ for $\omega_2$. Obviously, the frequencies do not depend on the wave-number when those 
relations between the model parameters are obeyed. Asymptotically the soliton approaches zero on one side
and $\pm1$ on the other. When scattering phonons off the soliton, the frequency bands for the reflected
and transmitted phonons are thus different.

The group velocity is defined as the derivative of the frequency with respect to the wave-number
\begin{equation}
\begin{cases}
v_{g_1}=\frac{\rm{d} \omega_{\mit{1}}}{\rm{d} \mit{q}}=\frac{h}{4}
\frac{(\frac{4}{h^2}-\frac{1}{2})\sin(qh)}{\sqrt{(\frac{4}{h^2}-\frac{1}{2})\sin^2 \frac{qh}{2}+\frac{1}{2}}} 
& \quad \mbox{for}\quad \phi_{n\pm 1}^0=\phi_n^0=0,\\
v_{g_2}=\frac{\rm{d} \omega_{\mit{2}}}{\rm{d} \mit{q}}
=\frac{h}{4}\frac{(\frac{4}{h^2}-2)\sin(qh)}{\sqrt{(\frac{4}{h^2}-2)\sin^2 \frac{qh}{2}+2}}
& \quad \mbox{for}\quad \phi_{n\pm 1}^0=\phi_n^0=\pm1.
\end{cases}
\label{eq:GroupVelocity}
\end{equation}
As is common for oscillations of systems with discretized translation invariance, the group velocity vanishes 
for $q\rightarrow 0$ and $q\rightarrow \pm\pi/h$ representing standing waves. The group velocity is maximal in 
the middle of the phonon band where the dispersion relation is steepest. 

\section{Numerical Results}\label{Phonon-kink}

\begin{figure*}[ht!]
\begin{center}
  \centering
  \subfigure[]{\includegraphics[width=0.40\textwidth,,height=0.15\textheight]
{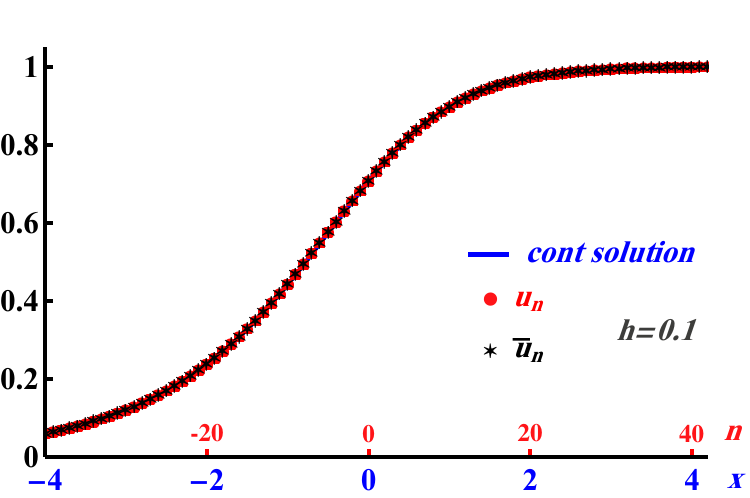}\label{fig:Phi6FieldAnalyticallyh01}}
 \hspace{5mm}
  \subfigure[]{\includegraphics[width=0.40\textwidth,height=0.15\textheight]
{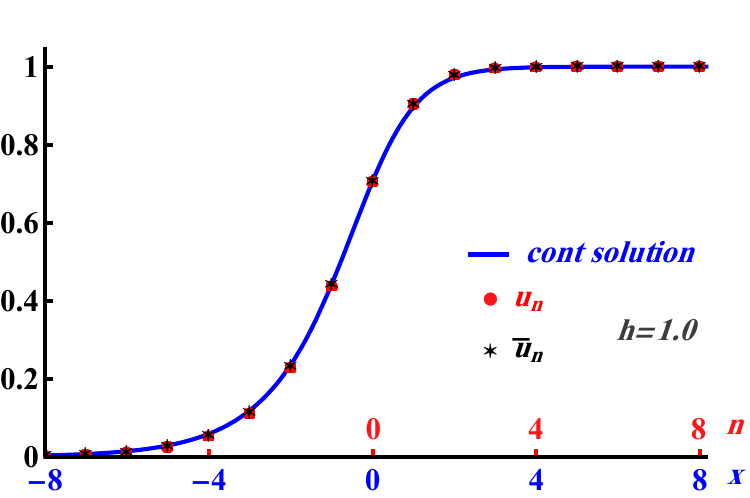}\label{fig:Phi6FieldAnalyticallyh10}}
\\
  \subfigure[]{\includegraphics[width=0.40\textwidth,height=0.15\textheight]
{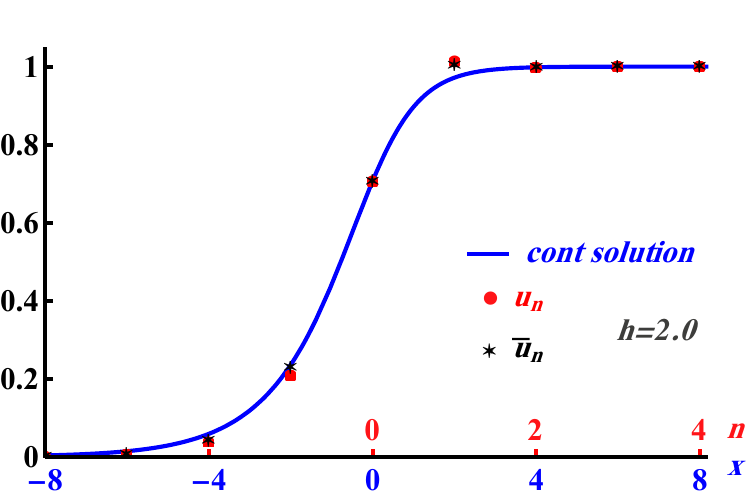}\label{fig:Phi6FieldAnalyticallyh20}}
  \hspace{5mm}
  \subfigure[]{\includegraphics[width=0.40\textwidth,height=0.15\textheight]
{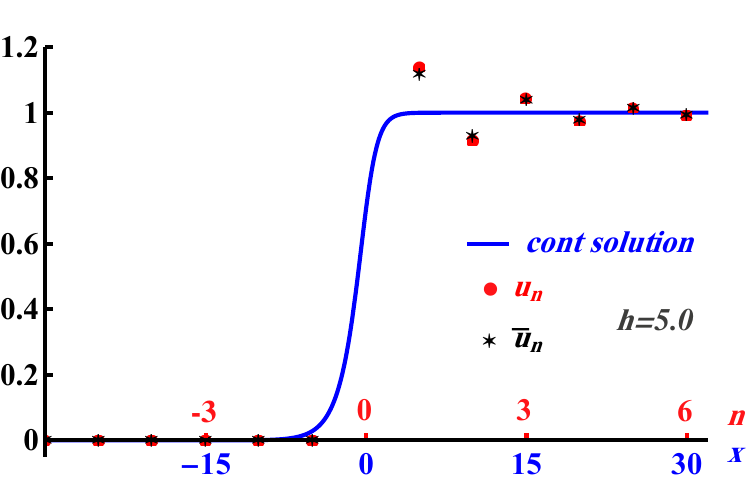}\label{fig:Phi6FieldAnalyticallyh50}}
 \caption{Comparison of the continuum $\phi^6$ kink solution (blue line) with the discrete kink profiles
for the discretizations given by Eq.~(\ref{eq:discrete1}) (red dots) and Eq.~(\ref{eq:NEW}) (black star) 
for different lattice spacings, $h$. Note the different scales for the horizontal axis.}
 \label{fig1}
\end{center}
\end{figure*}

To initialize the numerical analysis, we construct the static profiles from the cubic maps, 
Eqs.~(\ref{eq:discrete1}) and~(\ref{eq:NEW}). We have verified that these discretizations are indeed PNP free, 
{\it i.e.\@} we can prescribe any value between zero and one for $\phi_0$. Different choices yield identical 
kink-like structures that are related by continuous translations. The smaller $\phi_0$, the further to the right is 
the center of the structure. In Fig.~\ref{fig1} we display the numerical results for the choice 
$\phi_0=\phi_{(0,1)}(0)=1/\sqrt{2}$ listed in Eq.~(\ref{eq:Phi6Solution1}).
As expected, for small lattice spacings $h$ both discretizations essentially reproduce the continuum solution
$\phi_{(0,1)}$. As $h$ is increased, discrepancies emerge. Interestingly, the kink-like structures cease to 
be monotonic functions, rather they oscillate about the continuum solution. This behavior was also observed
in Ref.~\cite{Rakhmatullina2018-kz} for the prescription of Eq.~(\ref{eq:discrete1}). For larger lattice spacing, 
we indeed find (small) differences between the two discretizations with the BPS consistent prescription being 
somewhat closer to the continuum kink.

Exploring the interaction of a phonon wave-packet with a kink is the central object of this project.
This exploration follows a well-established procedure in kink dynamics: The interacting objects are 
components of a single field which can be disentangled asymptotically. The comparison of the individual
components at very early and very late times allows one to determine various scattering data. Even scattering 
phase shifts, that typically are extracted from small amplitude fluctuations about the kink, can be computed 
by this procedure \cite{Abdelhady.IJMPA.2011}. For a collection of numerous such calculations we refer to 
the recent review \cite{kevrekidis2019dynamical}.

With the choice that the kink profile vanishes at negative spatial infinity, the dispersion relation $\omega_1$
characterizes a phonon wave-packet to the left of the kink, while $\omega_2$ does so on the other side.
In order to excite a phonon wave-packet that eventually interacts with that kink, one site of the lattice, 
$n=n^{\ast}$ is dislocated. With the kink initially being at the center, $n=0$, an initial dislocation to 
its left or right would sit at $n^{\ast}\sim-\frac{N}{4}$ or $n^{\ast}\sim\frac{N}{4}$, respectively.
To generate a wave-packet peaked around a given frequency $\Omega$, we take at early times 
\begin{equation}
\epsilon_{n^{\ast}}(t)=A\sin(\Omega t)
\label{DrivingForce}
\end{equation}
and determine all other $\epsilon_n$ from the second order differential equation (\ref{eq:fluctuation})
with the initial conditions $\epsilon_n(0)=0$ and $\dot{\epsilon}_n(0)=0$ for $n\ne n^\ast$. At later times, 
but still long before the wave-packet reaches the boundary at $\pm N$ or the kink, all $\phi_n(t)$
are determined from Eq.~(\ref{eq:fieldequation}). Both the amplitude $A$ and the frequency $\Omega$ 
characterize the wave-packet, and we will explore its interaction with the kink as a function thereof.
The construction ensures that the wave-packet propagates symmetrically towards the boundary and the kink
as long as no interactions with either the boundary or the kink occur. This allows us to determine the
initial energy flowing towards the kink from the energy flowing towards the boundary.

We have simulated this interaction for $A=0.01$, $A=0.02$ and $A=0.05$. In what follows we will graphically
present and discuss the results specifically for $A=0.01$. There are no conceptual differences as long as the 
amplitude is consistent with the linearization of Eq.~(\ref{eq:fluctuation}).

\begin{figure*}[ht!]
\includegraphics[width=13.0cm,height=7cm]{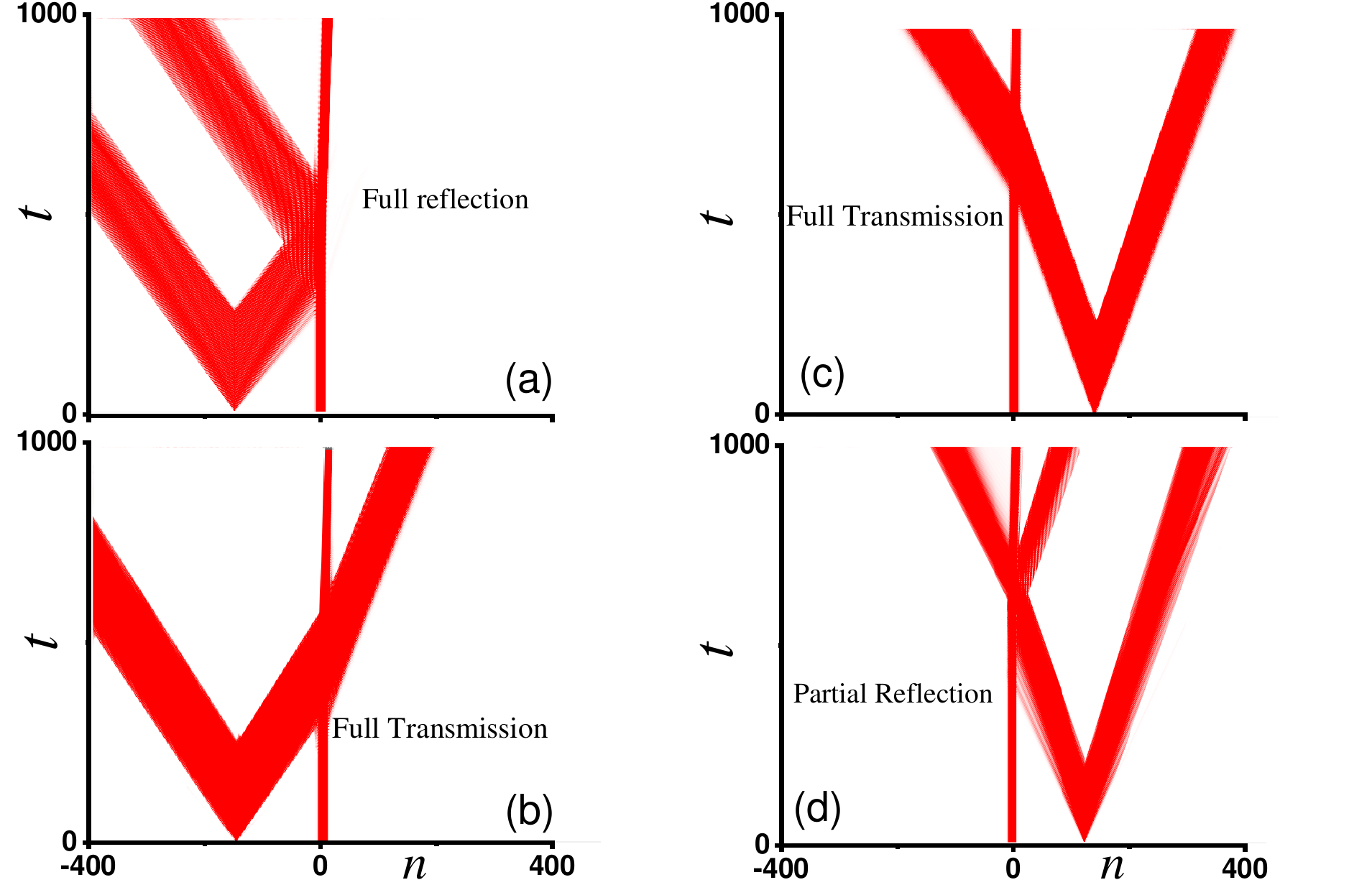}
\caption{Total energy flow (kinetic plus potential) of all particles and the energy source are shown 
for the case when phonons scatter from the left (a,c) or the right (b,d) side of the kink. 
The driving frequencies are (a) $\Omega=0.9$, (b) $\Omega=1.6$, (c) $\Omega=1.7$ and (d) $\Omega=1.9$. 
The kink is located at the center of the chain and the lattice spacing is $h=1.0$ in this case.} 
\label{EnergyFlow_h_1}
\end{figure*}

\begin{figure*}[ht!]
\includegraphics[width=13.0cm,height=7cm]{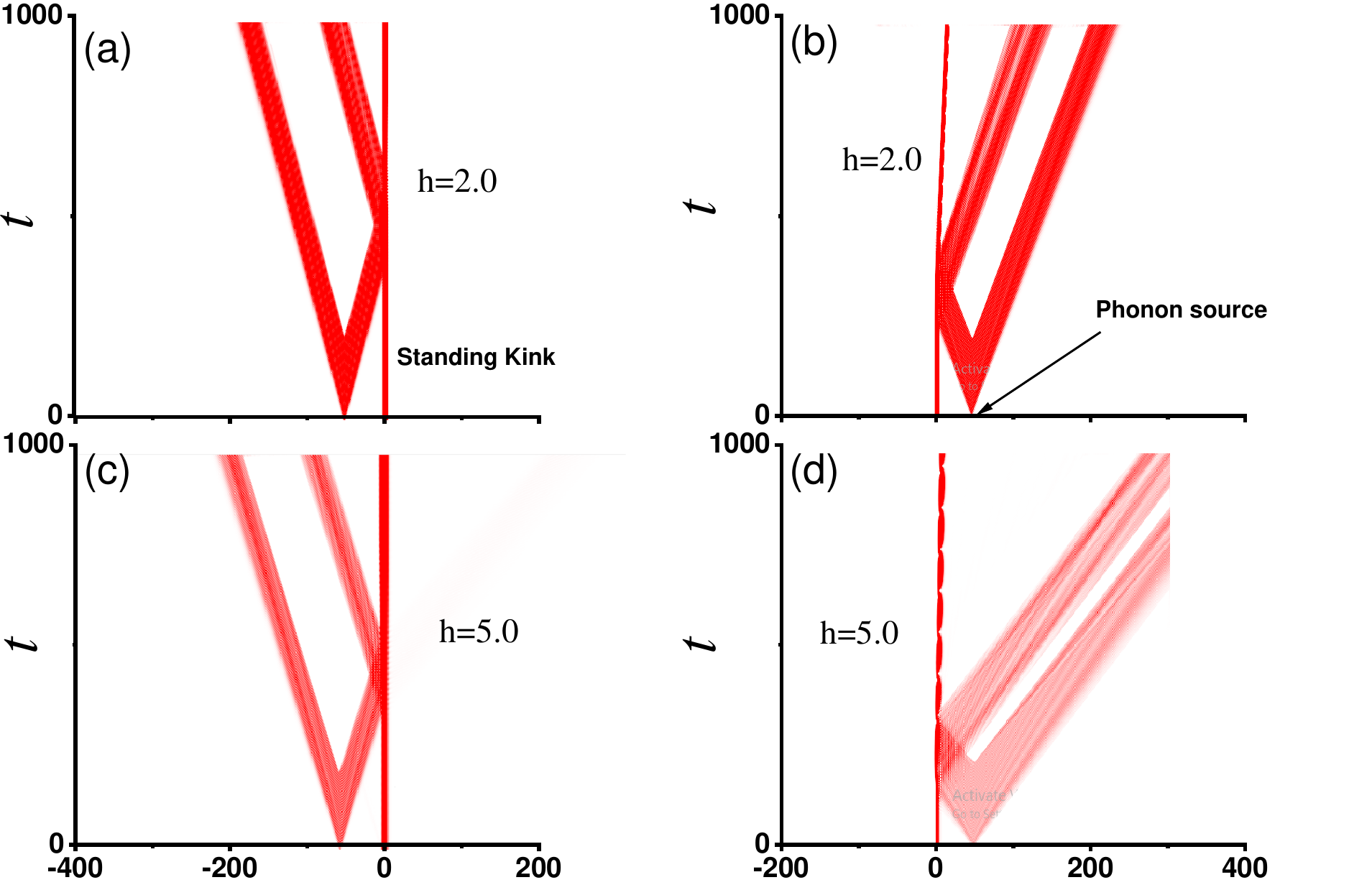}
\caption{Same as Fig.~\ref{EnergyFlow_h_1} with the wave-packet approaching the kink from the left and right for 
(a,b) and (c,d), respectively. The driving frequencies are (a) $\Omega=0.85$, (b) $\Omega=1.15$, (c) $\Omega=0.5$, 
and (d) $\Omega=1.1$.}
\label{EnergyFlow_h_2_h_5}
\end{figure*}

Let us discuss some sample simulations for the discretization prescription of Eq.~(\ref{eq:discrete1}).
In Fig.~\ref{EnergyFlow_h_1} we present the energy flow for a moderate lattice spacing, $h=1.0$. In that 
space-time plot, we indicate the regions in which the time dependent energy density, 
\begin{equation}\label{eq:EnergyPer}
e_n(t)=\frac{h}{2}\left(\dot{\phi}_n^2+u_n^2 \right), 
\end{equation}
is larger than a threshold value, $e_{\rm thr.}$ for which we take $e_{\rm thr.}=10^{-6}$.
For this value of $h$, a wave-packet approaching the kink from the left is either (almost) fully reflected or 
fully transmitted. Later we will see that there is a small window for the driving frequency in which partial 
reflection and transmission occur. An example is displayed in part d) of Fig.~\ref{EnergyFlow_h_1} for a 
wave-packet approaching the kink from the right. Fig.~\ref{EnergyFlow_h_2_h_5} suggests that
transmission disappears for larger lattice spacings.

\begin{figure*}[t!]
\begin{center}
  \centering
  \subfigure[]{\includegraphics[width=0.35
 \textwidth, height=0.16 \textheight]{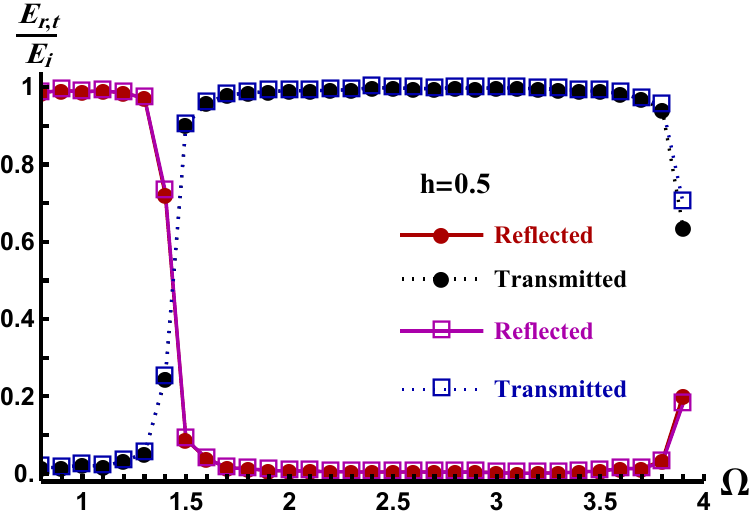}}
    \subfigure[]{\includegraphics[width=0.35
 \textwidth, height=0.16 \textheight]{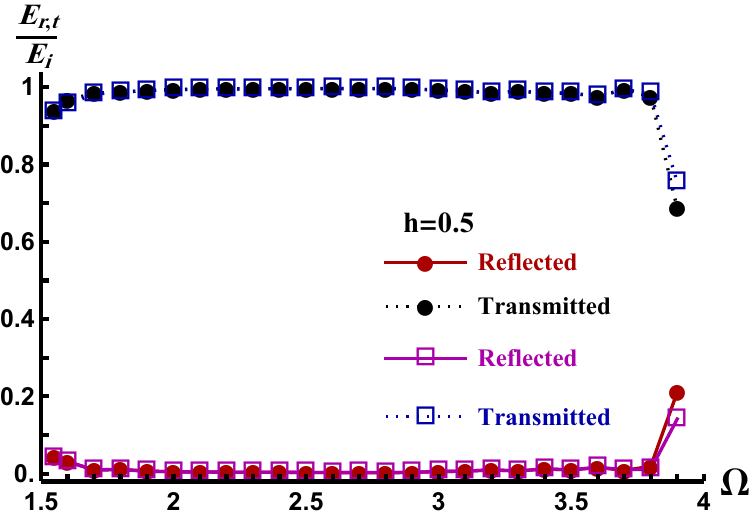}}
\\
  \subfigure[]{\includegraphics[width=0.35
 \textwidth, height=0.16 \textheight]{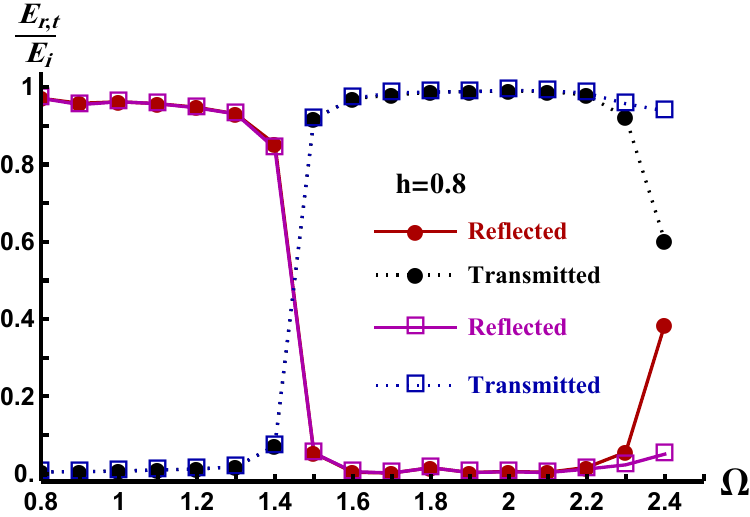}}
    \subfigure[]{\includegraphics[width=0.35
 \textwidth, height=0.16 \textheight]{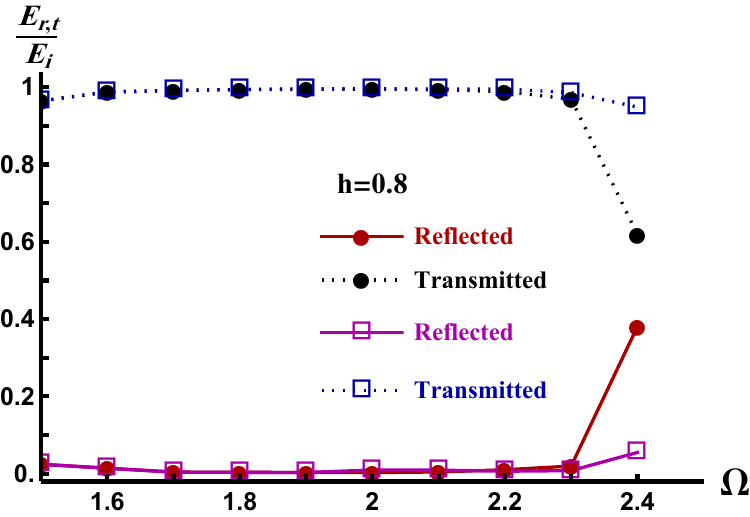}}
\\
  \subfigure[]{\includegraphics[width=0.35
 \textwidth, height=0.16 \textheight]{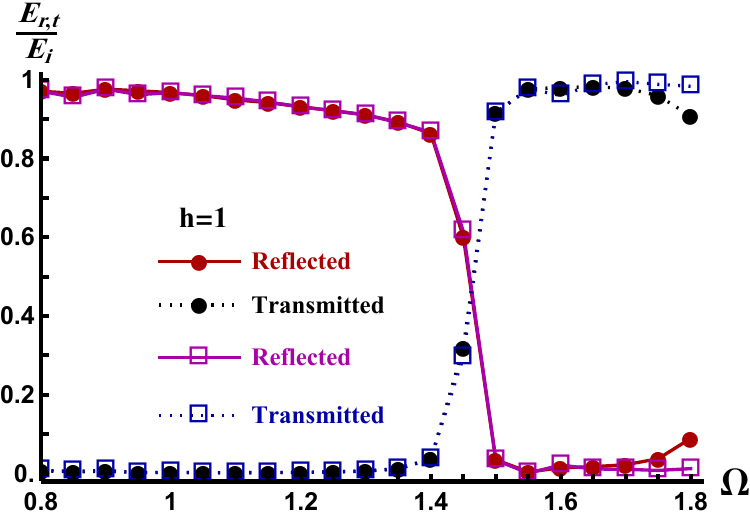}}
    \subfigure[]{\includegraphics[width=0.35
 \textwidth, height=0.16 \textheight]{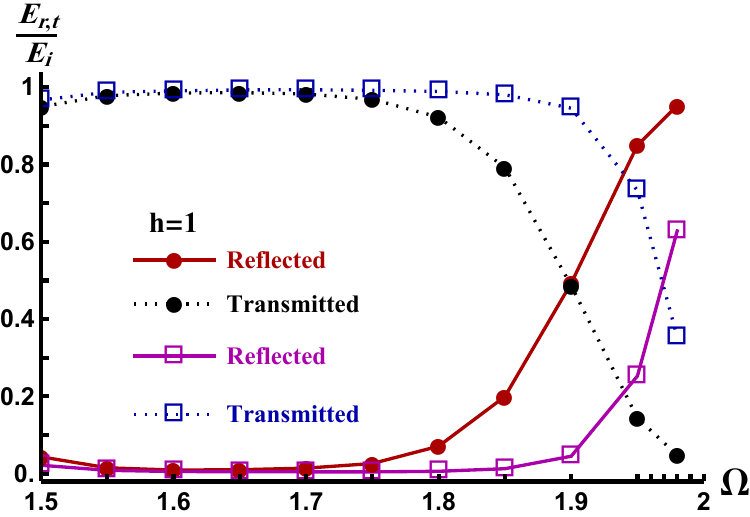}}
\\
  \subfigure[]
{\includegraphics[width=0.35\textwidth, height=0.16 \textheight]{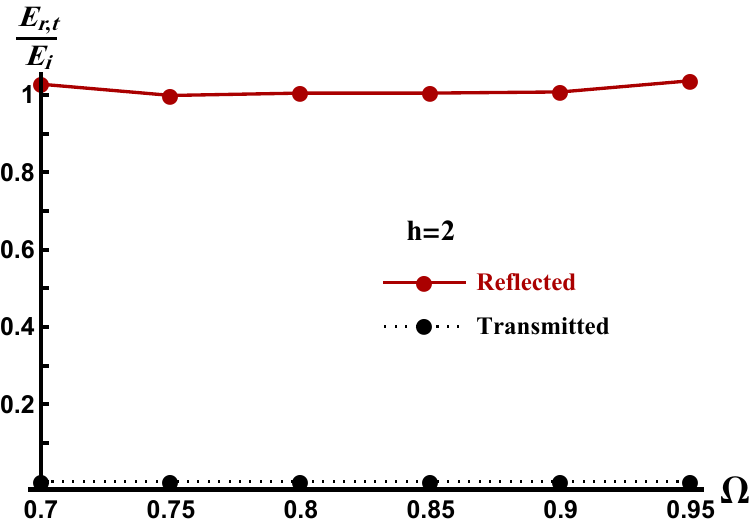}}
  \subfigure[]
{\includegraphics[width=0.35\textwidth, height=0.16 \textheight]{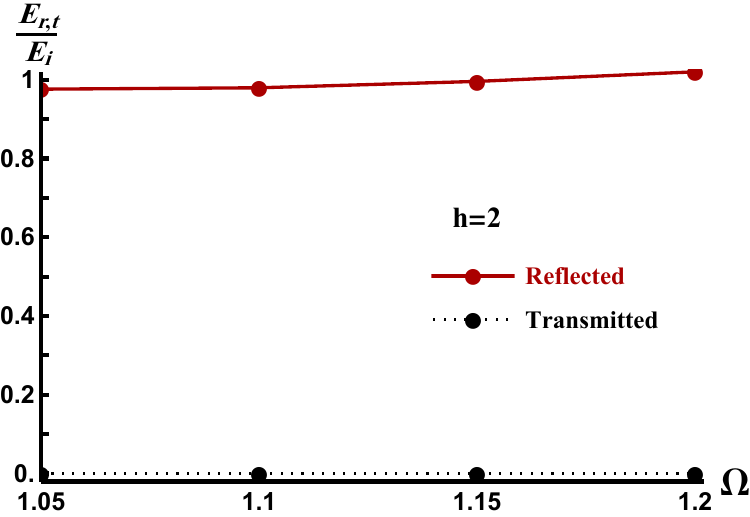}}
\\
\caption{Normalized transmission and reflection coefficients for phonons scattering off the kink as functions of the 
driving frequency $\Omega$ for different values of the lattice spacing $h$ as specified in the legends. Panels 
(a), (c), (e) and (g) refer to phonons originating from the left side of the kink with $E_r/E_i$ and $E_t/_i$
computed from Eq.~(\ref{eq:coefficients1}) while the entries (b), (d), (f) and (h) have the  phonons starting 
from the right side of the kink using Eq.~(\ref{eq:coefficients2}). In all cases, the entire range of the phonon 
band is displayed. Full circles and open squares refer to the discretization prescriptions from 
Eqs.~(\ref{eq:discrete1}) and~(\ref{eq:NEW}), respectively. Continuous or dotted lines have been included to guide the eye.}
  \label{fig:EnergyTrans}
\end{center}
\end{figure*}

To provide a more quantitative analysis, we explore the energy density in  Eq.~(\ref{eq:EnergyPer}) to define 
initial, reflected and transmitted energies as introduced in Ref.~\cite{Evazzade2018-fi}
\begin{equation}\label{eq:coefficients1}
E_i=\frac{1}{2}\sum_{n=-N}^{-n_k} e_n(t^\prime)\,,\quad  
E_r=\sum_{n=-N}^{-n_k} e_n(t^{\prime\prime}) -E_i  
\quad {\rm and}\quad 
E_t=\sum_{n=n_k}^{N}  e_n(t^{\prime\prime})\,,
\end{equation}
when the wave-packet originates from the left. We eliminate the kink contribution to the energies by setting 
$n_k\approx 5$.  Here, $t^\prime$ is any time after $\epsilon_{n^\ast}$ is no longer fixed by Eq.~(\ref{DrivingForce}) 
but before the interaction with the kink takes place. Furthermore, $t^{\prime\prime}$ denotes an instant past that 
interaction. 
For the phonons propagating from the opposite direction, the energies are 
\begin{equation}\label{eq:coefficients2}
E_i=\frac{1}{2}\sum_{n_k}^{N} e_n(t^\prime)\,,\quad  
E_r=\sum_{n=n_k}^{N} e_n(t^{\prime\prime}) -E_i\
\quad {\rm and}\quad 
E_t=\sum_{n=-n_k}^{-N}  e_n(t^{\prime\prime})\,.
\end{equation}
Note that $n^\ast$ is different for the scenarios to which  Eqs.~(\ref{eq:coefficients1}) and~(\ref{eq:coefficients2})
are applied. We have varied $t^\prime$ and $t^{\prime\prime}$ within sensible intervals and found stable results.
For the tiny amplitude of phonons considered here, almost no energy radiation from or displacement of the kink is 
observed. Consequently, $E_r\approx E_i-E_t$, which is substantiated by the results shown in Fig.~\ref{fig:EnergyTrans}.

Before discussing details of those, we observe that interactions only occur when the driving frequency~$\Omega$ 
is within the bands for $\omega$ shown in Fig.~\ref{fig:omegah} even though there is no direct relation between $\omega$ and 
$\Omega$. There is no impediment for taking $\Omega$ outside those bands, but we find that for such choices, the initial excitation 
does not propagate and never reaches the kink. This is consistent with the principles of lattice vibrations: for frequencies
outside the band(s) the equations for the small amplitude fluctuations do not have (real) solutions, and the lattice is
not able to accommodate oscillations with such frequencies.

In Fig.~\ref{fig:EnergyTrans} we present the transmission and reflection coefficients from Eqs.~(\ref{eq:coefficients1})
and~(\ref{eq:coefficients2}) as functions of the driving frequency $\Omega$ for different lattice spacings $h$. The entries 
in the left and right panels refer to the wave-packets approaching the kink from negative and positive infinity, respectively.
For small lattice spacings, the phonon wave-packet can be either reflected or transmitted by the kink. At low frequencies (energies)
and when the phonons approach the kink from its left hand side, the phonons cannot pass through the kink, while the 
kink is more transparent for phonons close to the upper edge of the spectrum. Surprisingly, the transition from strong reflection 
to strong transmission is quite sharp and happens at the transition frequency $\Omega_{\rm t}\approx 1.5$. The figures 
suggest that a reverse transition materializes at the upper end of the frequency band. For the wave-packet 
originating from the right, we observe a different picture for small lattice spacings: The smaller the frequency, the more 
dominating is transmission. For large lattice spacing only reflection occurs. The frequency bands become narrower as the
lattice spacing increases: for phonons originating at the left of the kink, the transition frequency leaves the band at 
the upper end, while it does so at the lower end for the case that the phonons approach the kink from the right.
Fig.~\ref{fig:EnergyTrans} clearly supports that conclusion for the case when the phonons scatter from the left hand 
side of the kink. For the other direction, it is not as obvious. Unfortunately, in this case the frequency band
is quite narrow in the transition region which prevents a detailed analysis.

Somewhat surprisingly, the two discretization methods yield different results only when the phonons approach the kink from the 
right and only for moderate lattice spacings $h$, even though the methods should agree when $h$ tends to zero. First, note that 
$h=1.0$ is not very small when compared to the size of the domain on which the soliton exhibits significant changes. Second, 
the differences only become visible for high frequencies that are not accessible for large lattice spacings.  However, for 
smaller spacings, agreement is restored.

The largest spacing shown in Fig.~\ref{fig:EnergyTrans} is $h=2.0$. We have also computed the reflection and transmission 
coefficients for $h>2.0$ and observed (total) reflection as indicated by the bottom entries in Fig.~\ref{EnergyFlow_h_2_h_5}.

\begin{figure*}[ht!]
\includegraphics[width=13.0cm,height=9cm]{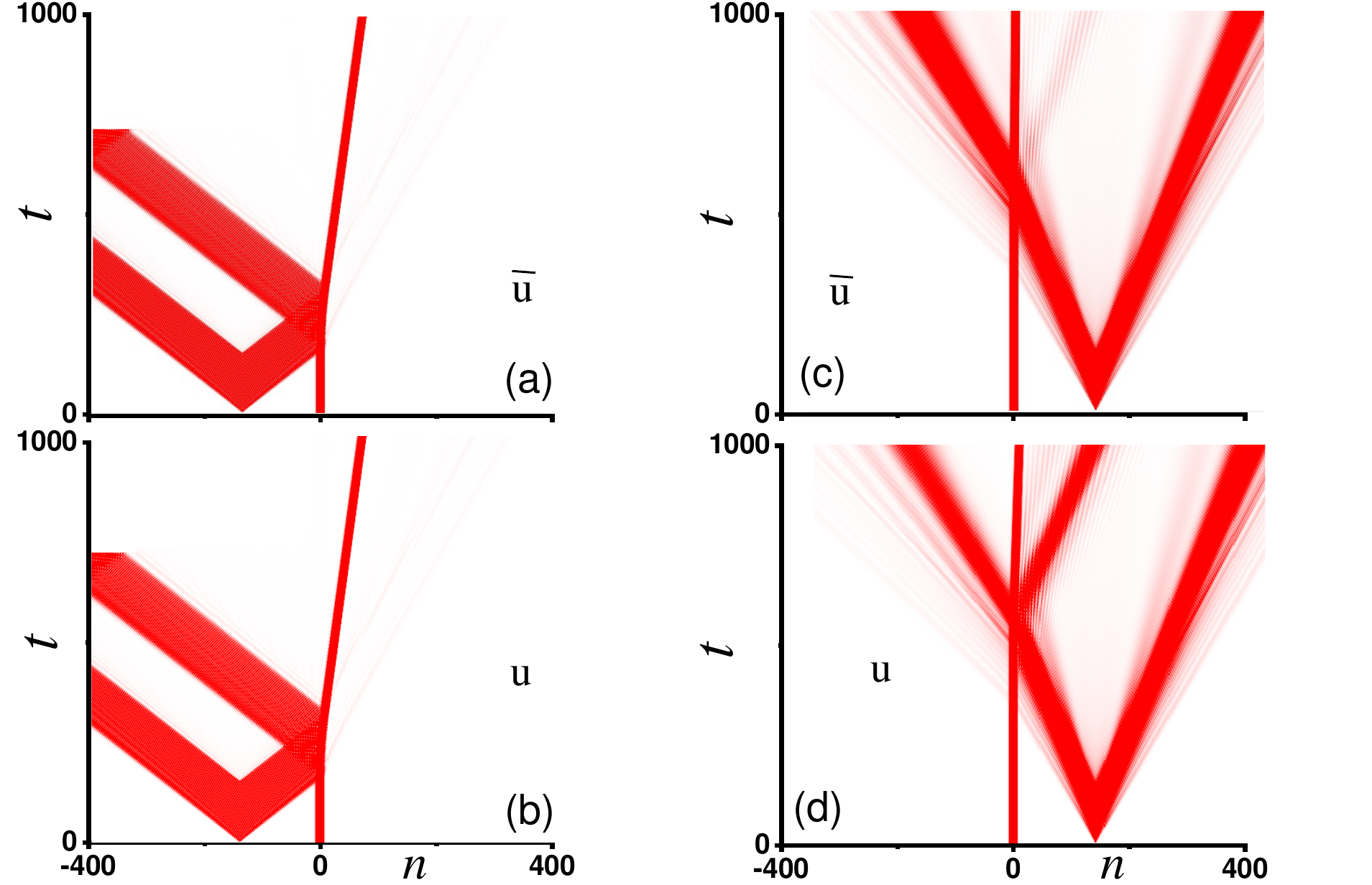}
\caption{Energy flow for cases when phonons scatter from the left (a,c) and right (b,d) of sides of the kink. 
The top and bottom entries are computed with the discretization prescriptions from Eqs.~(\ref{eq:discrete1}) 
and~(\ref{eq:NEW}), respectively. The driving frequencies are $\Omega=1.22$ in (a,c) and $\Omega=1.9$ in (b,d). Each
time the lattice spacing is $h=1.0$.}
\label{EnergyFlow_h_1_MoveKink}
\end{figure*}

A major motivation to study PNP free discretization prescriptions was the possibility to place the kink anywhere on the lattice 
and dislocate it by an amount independent of the lattice spacing. Consequently, it should be possible for the kink to acquire a 
small amount of kinetic energy from the interaction with the phonon wave-packet. Although the energy flow diagrams, 
Fig.~\ref{EnergyFlow_h_1} suggest otherwise, for small lattice spacings we do find cases in which the kink moves with a 
finite velocity after the interaction. Examples for this phenomenon are shown in Fig.~\ref{EnergyFlow_h_1_MoveKink}. 
The effect is more pronounced for phonons approaching the kink from the left, and in this case, the difference between 
the two discretization prescription is marginal. There is only a slight kink dislocation when the phonons start 
from the right side of the kink. In this case, we see again that the prescription from Eq.~(\ref{eq:NEW}) has more 
transmission for larger values of $\Omega$.

\begin{figure*}[ht!]
\includegraphics[width=11.0cm,height=7cm]{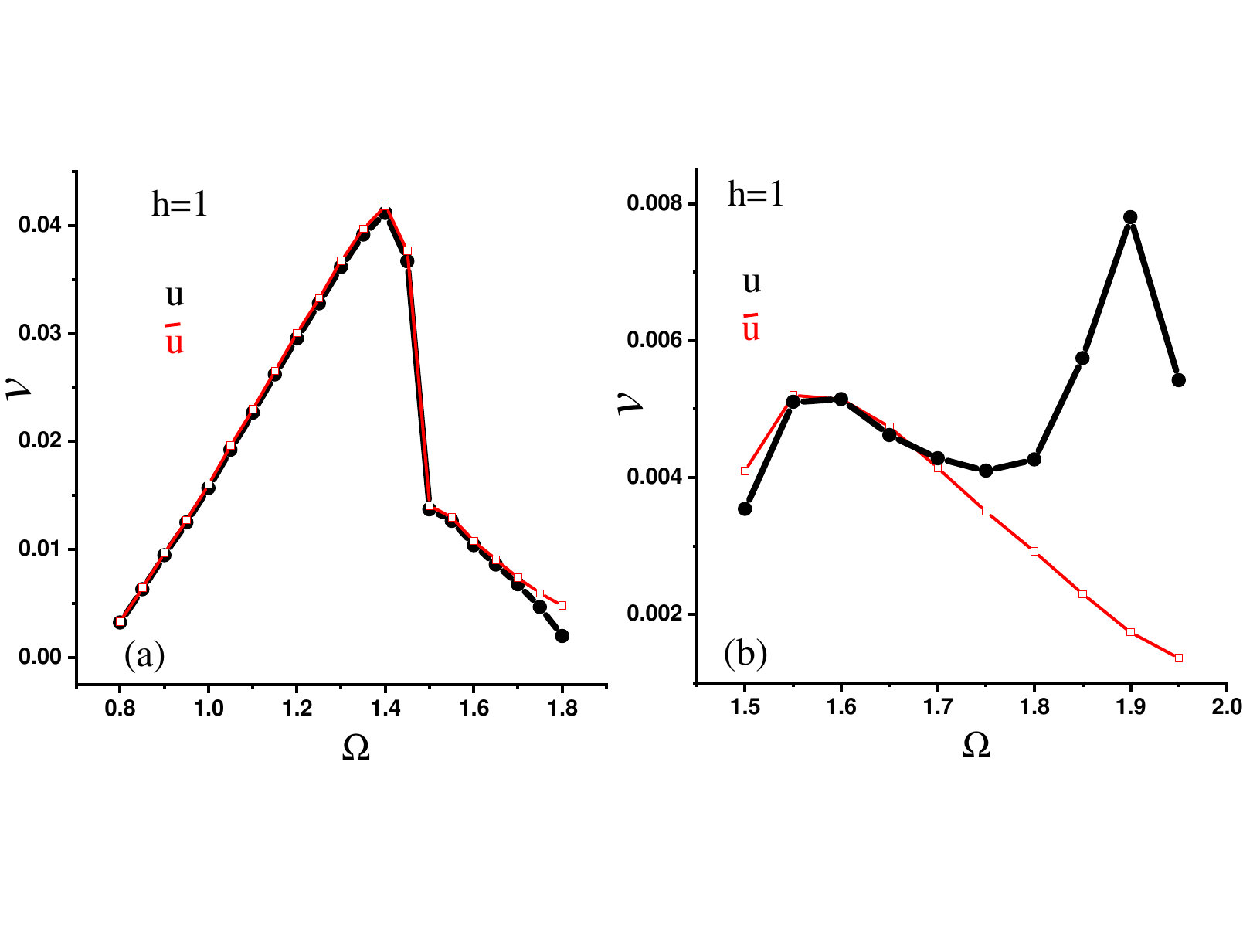}
\caption{Kink velocity as the function of the driving frequency $\Omega$ for the cases when phonons scatter (a) from the left 
and (b) the right sides of the kink for the both discretization prescriptions discussed in the text. The dimensionless 
lattice spacing is $h=1$. Note the different scales for the vertical axis.}
\label{Kink_Velocity}
\end{figure*}

We extract the velocity of the kink from the slope of its trajectory at late times. This velocity is shown in 
Fig.~\ref{Kink_Velocity} as a function of the driving frequency $\Omega$. 
This small velocity linearly increases with the frequency until the middle of the band is reached. Thereafter, it suddenly 
drops. Again, we see small differences between the two discretization methods at the end of the band. The figure suggests 
that the difference would be larger for the case with the kink coming from the right. However, in that case the velocity 
by itself is very small and the differences are actually similar in magnitude for the two methods. Interestingly enough, 
the kink always moves to the right regardless of whether the wave-packets originates from the left or the right. 

With our choice of the topological sector, a right-moving kink eventually causes the vacuum $\phi_0=1$ to disappear. 
This is a result of the classical back-reaction from the phonons on the kink which has also been observed in
the continuum limit~\cite{Romanczukiewicz:2017hdu}\footnote{Interestingly, quantum effects make the kink move 
in the opposite direction \cite{Weigel:2017iup} because fewer modes contribute to the vacuum polarization
energy when the lower frequency bound increases.}. This may also be an explanation for the stiffness of the
soliton when the wave-packet originates from the right. In that case, we add negative momentum to the system which is 
difficult for the soliton to absorb because it prefers positive momentum. 

The above numerical results have all been obtained using (unphysical) dimensionless coordinates such that the model parameters 
do not explicitly appear. However, any particular choice of model parameters (including the lattice spacing $h$) can be 
accommodated by multiplying appropriate powers\footnote{This power is most easily determined by noting that $v^2\sqrt{k}$ has 
unit mass dimension.} of $v^2\sqrt{k/2}$.

\section{Conclusions}\label{sec:conclusions} 

We have considered discretized versions of the $\phi^6$ model in $D=1+1$ dimensions. The field potential has degenerate vacua, 
which are connected by the kink soliton, with different curvatures so that the resulting solitons are asymmetric. In the 
continuum limit these curvatures are the masses of the small amplitude fluctuations. We have chosen the boundary conditions 
in such a way that the larger mass is at positive spatial infinity, {\it i.e.\@} to the right of the kink. Despite facing 
challenges on the quantum level due to the different masses, a number of interesting phenomena can be studied classically 
in this model. For example, this asymmetry causes the kink-antikink scattering to differ from the antikink-kink one 
\cite{Dorey:2011yw,Takyi:2016tnc}. One expects similar differences for the interaction of wave-packets that approach the 
kink from opposite sides. In this project, we have done so for a discretized version of the $\phi^6$ model because 
band-structure dispersion relations emerge for the modes that build the wave-packet. Generally, the lattice structure 
of the discretization could fix the kink location inducing irregular footprints for the scattering data. Fortunately,
there are certain discretization prescriptions that allow us to place the kink anywhere along the lattice. They are called 
Peierls-Nabarro free (potentials) and we have exploited two versions thereof. One was proposed in Ref.~\cite{Rakhmatullina2018-kz} 
and we developed a new one which makes closer contact with the continuum BPS construction. Since the two prescriptions coincide 
in the limit of zero lattice spacing, it is not surprising that the kink profiles only differ for large spacings, even though 
the differences are marginal. We have chosen the topological sector in such a way that the kink profile vanishes at negative 
spatial infinity and assumes the non-zero positive vacuum expectation value at the other end. Furthermore, we have 
encountered a new interesting feature of the Peierls-Nabarro free discretization prescriptions in the non-soliton case: 
When the lattice spacing, $h$, exceeds a critical value, the dispersion relations for the lattice vibrations turn from 
being acoustic to optic.

A propagating phonon wave-packet is induced by some kind of brute-force method that inscribes a driving frequency. 
Together with the standing kink this serves as initial condition when solving the full time-dependent wave-equation.
By selecting the starting point of the wave-packet, we consider the cases in which it approaches the kink from positive 
or negative spatial infinity separately. The phonons are partially reflected and/or transmitted, depending on $h$.
For large values of $h$, phonons are always reflected from the kink, regardless of the direction of motion. 
However, for moderate values of $h$, both transmission and reflection are observed. In cases with transmission and reflection, 
the former dominates for large frequencies and the latter for small ones when the phonons come from the left. For phonons 
scattering from the right, it is the other way round. 

These new features of the phonon-kink interaction might have many applications for topological states of buckled graphene 
membranes whose interactions with the lattice structure have recently been studied in Ref.~\cite{PhysRevB.103.224312}.
We conjecture that these states act as obstacles for optical phonons (at large values of $h$) but are more transparent 
for acoustic phonons (at moderate or small values of $h$).

Since the Peierls-Nabarro free set-up allows the kink to be placed anywhere along the lattice, the wave-packet should be 
capable of repelling or attracting the kink. Indeed, we observe that the kink is repelled when the phonons come from the 
left side. However, the effect is small, as measured by the velocity of the final kink configuration. Additionally, there is 
a marginal attraction when the phonons come from that side of the kink where the continuum fluctuations have the larger mass. 
Similar behaviors were also observed in the $\phi^4$ model, both for the discretized  \cite{PhysRevE.96.042109} and 
continuum \cite{Abdelhady.IJMPA.2011} versions.

For future investigations, it would be particularly interesting to consider the scattering of large amplitude waves from a 
symmetric or asymmetric soliton. Several new interesting phenomena, such as soliton acceleration, the creation of multiple 
soliton and anti-soliton pairs, and resonant energy pumping, could eventually be observed \cite{Evazzade2018-fi,Moradi.EPJB.2022}.

\section*{Acknowledgments}

D.\@ S.\@ expresses gratitude to Prof.\@ B.\@ Malomed and Prof.\@ S. V.\@ Dmitriev for engaging in a productive discussion 
that provided valuable insights or ideas on this work. A.\@ M.\@ M would like to extend his sincere gratitude to 
Prof.\@ S.\@ V.\@ Dmitriev for his invaluable guidance and unwavering support during his time in Ufa. 
Furthermore, A.\@ M.\@ M is deeply thankful to Islamic Azad University, Quchan Branch, for generously providing the grant 
that supported this work.  H.\@ W.\@ is supported in part by the National Research Foundation of
South Africa (NRF) by grant~150672.

\appendix
\section{Field Equations}
\label{appendix}
In this appendix, we present the field equations mentioned in the main body of the paper. First, we write out 
$U_n$ from Eq.~(\ref{eq:fieldequation}) to formulate the second order differential equation
\begin{align}
\label{eq:EquationOfMotion}
\ddot{\phi}_n&=\Delta_2\phi_n+\frac{1}{6\sqrt{2}h}\big[(\phi_{n-1}^3-\phi_{n+1}^3)-3\phi_n^2(\phi_{n-1}-\phi_{n+1})\big] 
\nonumber\\ &\hspace{1.33cm}
+\frac{1}{72}\big[-2(\phi_{n-1}^5+\phi_{n+1}^5)-8\phi_n(\phi_{n-1}^4+\phi_{n+1}^4)+(9-15\phi_n^2)(\phi_{n-1}^3+\phi_{n+1}^3) 
\nonumber\\ &\hspace{2.5cm}
+(24\phi_n-16\phi_n^3)(\phi_{n-1}^2+\phi_{n+1}^2)+(27\phi_n^2-10\phi_n^4-9)(\phi_{n-1}+\phi_{n+1}) 
\nonumber \\ &\hspace{2.5cm}
-6\phi_n^5+24\phi_n^3-18\phi_n\big],
\end{align}
where $\Delta_2\phi_n=\frac{1}{h^2}(\phi_{n-1} -2\phi_n+\phi_{n+1})$. Next, we list the coefficient functions appearing in the linearized form of Eq.~(\ref{eq:fluctuation}) with the convention that 
$\phi_n=\phi_n^{(0)}$
\begin{align}
V_{n}^{(\pm)}&=\frac{1}{h^2}+\frac{1}{2\sqrt{2}h}(\phi_{n\pm1}^{2}-\phi_{n}^{2})
\nonumber\\ &\hspace{1cm}
+\frac{1}{72}\big[-10\phi_{n\pm1}^{4}-32\phi_{n}\phi_{n\pm1}^{3}
+9(3-5\phi_n^{2})\phi_{n-1}^{2}+16 \phi_n (3-2\phi_{n}^{2})\phi_{n\pm1}
+27\phi_{n}^{2}-10\phi_n^{4}-9\big]\,,
\nonumber\\[1mm]
V_{n}^{(0)}=&-\frac{2}{h^2}-\frac{1}{\sqrt{2}h}\phi_n(\phi_{n-1}-\phi_{n+1})
\nonumber\\ &\hspace{1cm}
+\frac{1}{36}\big[-4(\phi_{n-1}^{4}+\phi_{n+1}^{4})-15\phi_n (\phi_{n-1}^{3}+\phi_{n+1}^{3})
+12(1-2\phi_{n}^{2})(\phi_{n-1}^{2}+\phi_{n+1}^{2})
\nonumber\\ &\hspace{2.5cm}
+\phi_n(27-20\phi_{n}^{2})(\phi_{n-1}+\phi_{n+1}) -15\phi_n^{4}+36\phi_{n}^{2}-9\big]\,.
\label{eq:Phi6Liniriazation}
\end{align}

The analog of Eq.~(\ref{eq:expandpot}) for the discretization of Eq.~(\ref{eq:NEW}) is
(omitting overbars for simplicity)
\begin{equation}
U_n=\sqrt{2}\left[\frac{1}{\sqrt{2}h}-\frac{1}{4}
+\frac{1}{8}\left(3\phi_n^2+2\phi_n\phi_{n-1}+\phi_{n-1}^2\right)\right]u_n
-\sqrt{2}\left[\frac{1}{\sqrt{2}h}+\frac{1}{4}
-\frac{1}{8}\left(3\phi_n^2+2\phi_{n+1}\phi_n+\phi_{n+1}^2\right)\right]u_{n+1}
\label{eq:expandpot1}
\end{equation}
from which we get the field equation
\begin{align}
\label{eq:EquationOfMotion1}
\ddot{\phi}_n&=\Delta_2\phi_n-\frac{3}{16}\bigg[\phi_{n}^5+\frac{5}{6}\phi_{n}^4(\phi_{n-1}+\phi_{n+1})
+\phi_{n}^3\left(\phi_{n-1}^2+\phi_{n+1}^2-\frac{8}{3}\right)  
\nonumber\\ & \hspace{2.3cm}
+\phi_{n}^2(\phi_{n-1}^3+\phi_{n+1}^3-2\phi_{n-1}-2\phi_{n+1})
+\phi_{n}\left(\frac{1}{2}\phi_{n-1}^4+\frac{1}{2}\phi_{n+1}^4-\frac{4}{3}\phi_{n-1}^2
-\frac{4}{3}\phi_{n+1}^2+\frac{4}{3}\right) 
\nonumber \\ & \hspace{2.3cm}
+\frac{\phi_{n-1}^5}{6}-\frac{2\phi_{n-1}^3}{3}+\frac{2\phi_{n-1}}{3}
+\frac{\phi_{n+1}^5}{6}-\frac{2\phi_{n+1}^3}{3}+\frac{2\phi_{n+1}}{3}\bigg]\,.
\end{align}
Observe that there is no contribution proportional to $\sqrt{k}$. Such a term would be linear in the 
discretized version of the derivative and its absence is reminiscent of the BPS construction. Finally the
linearized version of Eq.~(\ref{eq:EquationOfMotion1}) leads to the coefficient functions
\begin{align}
V_{n}^{(\pm)}&=\frac{1}{h^2}-\frac{1}{32}\big[
5\phi_n^4+12\phi_n^3\phi_{n\pm1}+18\phi_n^2\phi_{n\pm1}^2-12\phi_n^2+12\phi_n\phi_{n\pm1}^3
-16\phi_n\phi_{n\pm1} +5\phi_{n\pm1}^4-12\phi_{n\pm1}^2+4\big]
\nonumber \\
V_{n}^{(0)}&=\frac{-2}{h^2}-\frac{1}{32}\big[
30\phi_n^4+20\phi_n^3(\phi_{n-1}+\phi_{n+1})+6\phi_n^2(3\phi_{n-1}+3\phi_{n+1}-8)
\nonumber \\ & \hspace{1.8cm}
+12\phi_n(\phi_{n-1}^3+\phi_{n+1}^3-2\phi_{n-1}-2\phi_{n+1})
+3\phi_{n-1}^4+3\phi_{n+1}^4-8\phi_{n-1}^2-8\phi_{n+1}+8\big]\,.
\end{align}

\section{Remark on Dispersion Relation}
\label{appendixB}
In the discussion of Eq.\@ (\ref{eq:Phi6Spectrom}) we mentioned that the appearance of two optic branches at 
large lattice spacing was due to the discretization prescription that avoided the Peierls-Nabarro potential. 
In this appendix, we will corroborate that conjecture by considering a seemingly more natural prescription for a 
Klein-Gordon theory
\begin{equation}
\mathcal{L}=\frac{1}{2}\dot{\phi}^2-\frac{1}{2}\phi^{\prime2}-\frac{1}{4}\phi^2
\qquad \Longrightarrow\qquad
\ddot{\phi}-\phi^{\prime\prime}+\frac{1}{2}\phi=0\,.
\label{eq:KG1}
\end{equation}
The discretization leading to $V_n^{(0,\pm1)}[0]$ analog to the one in Eq.\@ (\ref{eq:Vasymp}) is
$$
\int dx\, \phi^2(x)\quad \longrightarrow\quad  h\sum_n\left(\frac{\phi_{n-1}+\phi_n}{2}\right)^2\,.
$$
It produces the dispersion relation
$$
\omega^2=\left(\frac{4}{h^2}-\frac{1}{2}\right)\sin^2\frac{qh}{2}+\frac{1}{2}\,,
$$
which is optic when $h>\sqrt{2}$.

A more suggestive discretization prescription would be
\begin{equation}
\int dx\, \phi^2(x)\quad \longrightarrow \quad h\sum_n \phi_n^2\,.
\label{eq:KG2}
\end{equation}
This leads to the dispersion relation
\begin{equation}
\omega^2=\frac{4}{h^2}\sin^2\frac{qh}{2}+\frac{1}{2}\,,
\label{eq:KG3}
\end{equation}
which, in the first Brillouin zone, increases with $|q|$ for any value of the lattice spacing, alike the acoustic 
branch of lattice vibrations. Extending the Lagrangian by higher powers of the field ($\phi^4$ and/or $\phi^6$) but 
discretizing them locally as in Eq.\@ (\ref{eq:KG2}) might change the additive constant in the dispersion relation 
but will not modify the coefficient of $\sin^2(qh/2)$.

\bibliographystyle{apsrev}
\bibliography{refs}

\begin{thebibliography}{45}
\expandafter\ifx\csname natexlab\endcsname\relax\def\natexlab#1{#1}\fi
\expandafter\ifx\csname bibnamefont\endcsname\relax
  \def\bibnamefont#1{#1}\fi
\expandafter\ifx\csname bibfnamefont\endcsname\relax
  \def\bibfnamefont#1{#1}\fi
\expandafter\ifx\csname citenamefont\endcsname\relax
  \def\citenamefont#1{#1}\fi
\expandafter\ifx\csname url\endcsname\relax
  \def\url#1{\texttt{#1}}\fi
\expandafter\ifx\csname urlprefix\endcsname\relax\def\urlprefix{URL }\fi
\providecommand{\bibinfo}[2]{#2}
\providecommand{\eprint}[2][]{\url{#2}}

\bibitem[{\citenamefont{Lohe}(1979)}]{Lohe:1979mh}
\bibinfo{author}{\bibfnamefont{M.~A.} \bibnamefont{Lohe}},
  \bibinfo{journal}{Phys. Rev. D} \textbf{\bibinfo{volume}{20}},
  \bibinfo{pages}{3120} (\bibinfo{year}{1979}).

\bibitem[{\citenamefont{Dorey et~al.}(2011)\citenamefont{Dorey, Mersh,
  Romanczukiewicz, and Shnir}}]{Dorey:2011yw}
\bibinfo{author}{\bibfnamefont{P.}~\bibnamefont{Dorey}},
  \bibinfo{author}{\bibfnamefont{K.}~\bibnamefont{Mersh}},
  \bibinfo{author}{\bibfnamefont{T.}~\bibnamefont{Romanczukiewicz}},
  \bibnamefont{and} \bibinfo{author}{\bibfnamefont{Y.}~\bibnamefont{Shnir}},
  \bibinfo{journal}{Phys. Rev. Lett.} \textbf{\bibinfo{volume}{107}},
  \bibinfo{pages}{091602} (\bibinfo{year}{2011}).

\bibitem[{\citenamefont{Nambu}(1977)}]{Nambu:1977ag}
\bibinfo{author}{\bibfnamefont{Y.}~\bibnamefont{Nambu}},
  \bibinfo{journal}{Nucl. Phys. B} \textbf{\bibinfo{volume}{130}},
  \bibinfo{pages}{505} (\bibinfo{year}{1977}).

\bibitem[{\citenamefont{Weigel}(2008)}]{Weigel:2008zz}
\bibinfo{author}{\bibfnamefont{H.}~\bibnamefont{Weigel}},
  \emph{\bibinfo{title}{{Chiral Soliton Models for Baryons}}}, vol.
  \bibinfo{volume}{743} (\bibinfo{publisher}{Lect. Notes Phys},
  \bibinfo{year}{2008}).

\bibitem[{\citenamefont{Feist et~al.}(2013)\citenamefont{Feist, Lau, and
  Manton}}]{Feist:2012ps}
\bibinfo{author}{\bibfnamefont{D.~T.~J.} \bibnamefont{Feist}},
  \bibinfo{author}{\bibfnamefont{P.~H.~C.} \bibnamefont{Lau}},
  \bibnamefont{and} \bibinfo{author}{\bibfnamefont{N.~S.}
  \bibnamefont{Manton}}, \bibinfo{journal}{Phys. Rev. D}
  \textbf{\bibinfo{volume}{87}}, \bibinfo{pages}{085034}
  (\bibinfo{year}{2013}).

\bibitem[{\citenamefont{{U. Schollw\"{o}ck, {\it et
  al.}}}(2004)}]{Schollwock:2004aa}
\bibinfo{author}{\bibnamefont{{U. Schollw\"{o}ck, {\it et al.}}}},
  \emph{\bibinfo{title}{{Quantum Magnetism}}}, vol. \bibinfo{volume}{645}
  (\bibinfo{publisher}{Lect. Notes Phys}, \bibinfo{year}{2004}).

\bibitem[{\citenamefont{Nagaosa and Tokura}(2013)}]{Nagasoa:2013}
\bibinfo{author}{\bibfnamefont{N.}~\bibnamefont{Nagaosa}} \bibnamefont{and}
  \bibinfo{author}{\bibfnamefont{Y.}~\bibnamefont{Tokura}},
  \bibinfo{journal}{Nature Nanotech.} \textbf{\bibinfo{volume}{8}},
  \bibinfo{pages}{899} (\bibinfo{year}{2013}).

\bibitem[{\citenamefont{Vilenkin and Shellard}(2000)}]{Vilenkin:2000jqa}
\bibinfo{author}{\bibfnamefont{A.}~\bibnamefont{Vilenkin}} \bibnamefont{and}
  \bibinfo{author}{\bibfnamefont{E.~P.~S.} \bibnamefont{Shellard}},
  \emph{\bibinfo{title}{{Cosmic Strings and Other Topological Defects}}}
  (\bibinfo{publisher}{Cambridge University Press}, \bibinfo{year}{2000}).

\bibitem[{\citenamefont{Moradi~Marjaneh
  et~al.}(2022)\citenamefont{Moradi~Marjaneh, Simas, and
  Bazeia}}]{MoradiMarjaneh:2022vov}
\bibinfo{author}{\bibfnamefont{A.}~\bibnamefont{Moradi~Marjaneh}},
  \bibinfo{author}{\bibfnamefont{F.~C.} \bibnamefont{Simas}}, \bibnamefont{and}
  \bibinfo{author}{\bibfnamefont{D.}~\bibnamefont{Bazeia}},
  \bibinfo{journal}{{Chaos Solitons and Fractals}}
  \textbf{\bibinfo{volume}{164}}, \bibinfo{pages}{112723}
  (\bibinfo{year}{2022}).

\bibitem[{\citenamefont{Nguyen et~al.}(2021)\citenamefont{Nguyen, Yamaletdinov,
  and Pershin}}]{PhysRevB.103.224312}
\bibinfo{author}{\bibfnamefont{D.~C.} \bibnamefont{Nguyen}},
  \bibinfo{author}{\bibfnamefont{R.~D.} \bibnamefont{Yamaletdinov}},
  \bibnamefont{and} \bibinfo{author}{\bibfnamefont{Y.~V.}
  \bibnamefont{Pershin}}, \bibinfo{journal}{Phys. Rev. B}
  \textbf{\bibinfo{volume}{103}}, \bibinfo{pages}{224312}
  (\bibinfo{year}{2021}).

\bibitem[{\citenamefont{Javidan}(2008)}]{Javidan.PRE.2008}
\bibinfo{author}{\bibfnamefont{K.}~\bibnamefont{Javidan}},
  \bibinfo{journal}{Phys. Rev. E} \textbf{\bibinfo{volume}{78}},
  \bibinfo{pages}{046607} (\bibinfo{year}{2008}).

\bibitem[{\citenamefont{Abdelhady and Weigel}(2011)}]{Abdelhady.IJMPA.2011}
\bibinfo{author}{\bibfnamefont{A.~M.~H.~H.} \bibnamefont{Abdelhady}}
  \bibnamefont{and} \bibinfo{author}{\bibfnamefont{H.}~\bibnamefont{Weigel}},
  \bibinfo{journal}{Int. J. Mod. Phys. A} \textbf{\bibinfo{volume}{26}},
  \bibinfo{pages}{3625} (\bibinfo{year}{2011}).

\bibitem[{\citenamefont{Bai et~al.}(2016)\citenamefont{Bai, Malomed, and
  Deng}}]{PhysRevE.94.032216}
\bibinfo{author}{\bibfnamefont{X.-D.} \bibnamefont{Bai}},
  \bibinfo{author}{\bibfnamefont{B.~A.} \bibnamefont{Malomed}},
  \bibnamefont{and} \bibinfo{author}{\bibfnamefont{F.-G.} \bibnamefont{Deng}},
  \bibinfo{journal}{Phys. Rev. E} \textbf{\bibinfo{volume}{94}},
  \bibinfo{pages}{032216} (\bibinfo{year}{2016}).

\bibitem[{\citenamefont{Campos and Mohammadi}(2022)}]{Azadeh.JHEP.2022}
\bibinfo{author}{\bibfnamefont{J.}~\bibnamefont{Campos}} \bibnamefont{and}
  \bibinfo{author}{\bibfnamefont{A.}~\bibnamefont{Mohammadi}},
  \bibinfo{journal}{JHEP} \textbf{\bibinfo{volume}{8}}, \bibinfo{pages}{180}
  (\bibinfo{year}{2022}).

\bibitem[{\citenamefont{{A.~Alonso-Izquierdo, D.~Miguélez-Caballero,
  L.~M.~Nieto, and J.~Queiroga-Nunes}}(2023)}]{Alonso.PhysD.2023}
\bibinfo{author}{\bibnamefont{{A.~Alonso-Izquierdo, D.~Miguélez-Caballero,
  L.~M.~Nieto, and J.~Queiroga-Nunes}}}, \bibinfo{journal}{Physica. D}
  \textbf{\bibinfo{volume}{443}}, \bibinfo{pages}{133590}
  (\bibinfo{year}{2023}).

\bibitem[{\citenamefont{Belova and Kudryavtsev}(1997)}]{Belova:1997bq}
\bibinfo{author}{\bibfnamefont{T.~I.} \bibnamefont{Belova}} \bibnamefont{and}
  \bibinfo{author}{\bibfnamefont{A.~E.} \bibnamefont{Kudryavtsev}},
  \bibinfo{journal}{Phys. Usp.} \textbf{\bibinfo{volume}{40}},
  \bibinfo{pages}{359} (\bibinfo{year}{1997}).

\bibitem[{\citenamefont{Kevrekidis and (eds.)}(2019)}]{Kevrekidis:2019}
\bibinfo{author}{\bibfnamefont{P.~G.} \bibnamefont{Kevrekidis}}
  \bibnamefont{and} \bibinfo{author}{\bibfnamefont{J.~C.-M.}
  \bibnamefont{(eds.)}}, \emph{\bibinfo{title}{{A dynamical perspective on the
  $\phi^4$ model : past, present and future}}}, Nonlinear systems and
  complexity, volume 26 (\bibinfo{publisher}{Springer}, \bibinfo{address}{Cham,
  Switzerland}, \bibinfo{year}{2019}), ISBN \bibinfo{isbn}{9783030118396}.

\bibitem[{\citenamefont{Braun and Kivshar}(2010)}]{Braun2010-ux}
\bibinfo{author}{\bibfnamefont{O.~M.} \bibnamefont{Braun}} \bibnamefont{and}
  \bibinfo{author}{\bibfnamefont{Y.~S.} \bibnamefont{Kivshar}},
  \emph{\bibinfo{title}{The Frenkel-Kontorova model}}, Theoretical and
  Mathematical Physics (\bibinfo{publisher}{Springer},
  \bibinfo{address}{Berlin, Germany}, \bibinfo{year}{2010}), ISBN
  \bibinfo{isbn}{{978-3540407713}}.

\bibitem[{\citenamefont{Bishop et~al.}(1980)\citenamefont{Bishop, Krumhansl,
  and Trullinger}}]{BISHOP19801}
\bibinfo{author}{\bibfnamefont{A.}~\bibnamefont{Bishop}},
  \bibinfo{author}{\bibfnamefont{J.}~\bibnamefont{Krumhansl}},
  \bibnamefont{and}
  \bibinfo{author}{\bibfnamefont{S.}~\bibnamefont{Trullinger}},
  \bibinfo{journal}{Physica D: Nonlinear Phenomena}
  \textbf{\bibinfo{volume}{1}}, \bibinfo{pages}{1} (\bibinfo{year}{1980}).

\bibitem[{\citenamefont{Derlet et~al.}(2007)\citenamefont{Derlet, Nguyen-Manh,
  and Dudarev}}]{PhysRevB.76.054107}
\bibinfo{author}{\bibfnamefont{P.~M.} \bibnamefont{Derlet}},
  \bibinfo{author}{\bibfnamefont{D.}~\bibnamefont{Nguyen-Manh}},
  \bibnamefont{and} \bibinfo{author}{\bibfnamefont{S.~L.}
  \bibnamefont{Dudarev}}, \bibinfo{journal}{Phys. Rev. B}
  \textbf{\bibinfo{volume}{76}}, \bibinfo{pages}{054107}
  (\bibinfo{year}{2007}).

\bibitem[{\citenamefont{Speight}(1999)}]{Speight:1998uq}
\bibinfo{author}{\bibfnamefont{J.~M.} \bibnamefont{Speight}},
  \bibinfo{journal}{Nonlinearity} \textbf{\bibinfo{volume}{12}},
  \bibinfo{pages}{1373} (\bibinfo{year}{1999}).

\bibitem[{\citenamefont{Barashenkov et~al.}(2005)\citenamefont{Barashenkov,
  Oxtoby, and Pelinovsky}}]{Barashenkov_PRE}
\bibinfo{author}{\bibfnamefont{I.~V.} \bibnamefont{Barashenkov}},
  \bibinfo{author}{\bibfnamefont{O.~F.} \bibnamefont{Oxtoby}},
  \bibnamefont{and} \bibinfo{author}{\bibfnamefont{D.~E.}
  \bibnamefont{Pelinovsky}}, \bibinfo{journal}{Phys. Rev. E}
  \textbf{\bibinfo{volume}{72}}, \bibinfo{pages}{035602}
  (\bibinfo{year}{2005}).

\bibitem[{\citenamefont{Dmitriev et~al.}(2006)\citenamefont{Dmitriev,
  Kevrekidis, Yoshikawa, and Frantzeskakis}}]{Dmitriev:2006qm}
\bibinfo{author}{\bibfnamefont{S.~V.} \bibnamefont{Dmitriev}},
  \bibinfo{author}{\bibfnamefont{P.~G.} \bibnamefont{Kevrekidis}},
  \bibinfo{author}{\bibfnamefont{N.}~\bibnamefont{Yoshikawa}},
  \bibnamefont{and} \bibinfo{author}{\bibfnamefont{D.~J.}
  \bibnamefont{Frantzeskakis}}, \bibinfo{journal}{Phys. Rev. E}
  \textbf{\bibinfo{volume}{74}}, \bibinfo{pages}{046609}
  (\bibinfo{year}{2006}).

\bibitem[{\citenamefont{Rakhmatullina et~al.}(2018)\citenamefont{Rakhmatullina,
  Kevrekidis, and Dmitriev}}]{Rakhmatullina2018-kz}
\bibinfo{author}{\bibfnamefont{Z.~G.} \bibnamefont{Rakhmatullina}},
  \bibinfo{author}{\bibfnamefont{P.~G.} \bibnamefont{Kevrekidis}},
  \bibnamefont{and} \bibinfo{author}{\bibfnamefont{S.~V.}
  \bibnamefont{Dmitriev}}, \bibinfo{journal}{IOP Conf. Ser. Mater. Sci. Eng.}
  \textbf{\bibinfo{volume}{447}}, \bibinfo{pages}{012057}
  (\bibinfo{year}{2018}).

\bibitem[{\citenamefont{Bogomolny}(1976)}]{Bogomolny:1975de}
\bibinfo{author}{\bibfnamefont{E.~B.} \bibnamefont{Bogomolny}},
  \bibinfo{journal}{Sov. J. Nucl. Phys.} \textbf{\bibinfo{volume}{24}},
  \bibinfo{pages}{449} (\bibinfo{year}{1976}).

\bibitem[{\citenamefont{Prasad and Sommerfield}(1975)}]{Prasad:1975kr}
\bibinfo{author}{\bibfnamefont{M.~K.} \bibnamefont{Prasad}} \bibnamefont{and}
  \bibinfo{author}{\bibfnamefont{C.~M.} \bibnamefont{Sommerfield}},
  \bibinfo{journal}{Phys. Rev. Lett.} \textbf{\bibinfo{volume}{35}},
  \bibinfo{pages}{760} (\bibinfo{year}{1975}).

\bibitem[{\citenamefont{Adam et~al.}(2019)\citenamefont{Adam, Oles, Queiruga,
  Romanczukiewicz, and Wereszczynski}}]{Adam:2019djg}
\bibinfo{author}{\bibfnamefont{C.}~\bibnamefont{Adam}},
  \bibinfo{author}{\bibfnamefont{K.}~\bibnamefont{Oles}},
  \bibinfo{author}{\bibfnamefont{J.~M.} \bibnamefont{Queiruga}},
  \bibinfo{author}{\bibfnamefont{T.}~\bibnamefont{Romanczukiewicz}},
  \bibnamefont{and}
  \bibinfo{author}{\bibfnamefont{A.}~\bibnamefont{Wereszczynski}},
  \bibinfo{journal}{JHEP} \textbf{\bibinfo{volume}{07}}, \bibinfo{pages}{150}
  (\bibinfo{year}{2019}).

\bibitem[{\citenamefont{Evslin et~al.}(2022)\citenamefont{Evslin, Halcrow,
  Romanczukiewicz, and Wereszczynski}}]{Evslin:2022xmp}
\bibinfo{author}{\bibfnamefont{J.}~\bibnamefont{Evslin}},
  \bibinfo{author}{\bibfnamefont{C.}~\bibnamefont{Halcrow}},
  \bibinfo{author}{\bibfnamefont{T.}~\bibnamefont{Romanczukiewicz}},
  \bibnamefont{and}
  \bibinfo{author}{\bibfnamefont{A.}~\bibnamefont{Wereszczynski}},
  \bibinfo{journal}{Phys. Rev. D} \textbf{\bibinfo{volume}{105}},
  \bibinfo{pages}{125002} (\bibinfo{year}{2022}).

\bibitem[{\citenamefont{Takyi and Weigel}(2023)}]{Takyi:2022sng}
\bibinfo{author}{\bibfnamefont{I.}~\bibnamefont{Takyi}} \bibnamefont{and}
  \bibinfo{author}{\bibfnamefont{H.}~\bibnamefont{Weigel}},
  \bibinfo{journal}{Phys. Rev. D} \textbf{\bibinfo{volume}{107}},
  \bibinfo{pages}{036003} (\bibinfo{year}{2023}).

\bibitem[{\citenamefont{Graham and Weigel}(2022)}]{Graham:2022rqk}
\bibinfo{author}{\bibfnamefont{N.}~\bibnamefont{Graham}} \bibnamefont{and}
  \bibinfo{author}{\bibfnamefont{H.}~\bibnamefont{Weigel}},
  \bibinfo{journal}{Int. J. Mod. Phys. A} \textbf{\bibinfo{volume}{37}},
  \bibinfo{pages}{2241004} (\bibinfo{year}{2022}).

\bibitem[{\citenamefont{Saadatmand and Marjaneh}(2022)}]{Moradi.EPJB.2022}
\bibinfo{author}{\bibfnamefont{D.}~\bibnamefont{Saadatmand}} \bibnamefont{and}
  \bibinfo{author}{\bibfnamefont{A.~M.} \bibnamefont{Marjaneh}},
  \bibinfo{journal}{Eur. Phys. J. B} \textbf{\bibinfo{volume}{95}},
  \bibinfo{pages}{144} (\bibinfo{year}{2022}).

\bibitem[{\citenamefont{Roma\'nczukiewicz}(2017)}]{Romanczukiewicz:2017hdu}
\bibinfo{author}{\bibfnamefont{T.}~\bibnamefont{Roma\'nczukiewicz}},
  \bibinfo{journal}{Phys. Lett. B} \textbf{\bibinfo{volume}{773}},
  \bibinfo{pages}{295} (\bibinfo{year}{2017}).

\bibitem[{\citenamefont{Currie et~al.}(1980)\citenamefont{Currie, Krumhansl,
  Bishop, and Trullinger}}]{PhysRevB.22.477}
\bibinfo{author}{\bibfnamefont{J.~F.} \bibnamefont{Currie}},
  \bibinfo{author}{\bibfnamefont{J.~A.} \bibnamefont{Krumhansl}},
  \bibinfo{author}{\bibfnamefont{A.~R.} \bibnamefont{Bishop}},
  \bibnamefont{and} \bibinfo{author}{\bibfnamefont{S.~E.}
  \bibnamefont{Trullinger}}, \bibinfo{journal}{Phys. Rev. B}
  \textbf{\bibinfo{volume}{22}}, \bibinfo{pages}{477} (\bibinfo{year}{1980}).

\bibitem[{\citenamefont{Graham et~al.}(2009)\citenamefont{Graham, Quandt, and
  Weigel}}]{Graham:2009zz}
\bibinfo{author}{\bibfnamefont{N.}~\bibnamefont{Graham}},
  \bibinfo{author}{\bibfnamefont{M.}~\bibnamefont{Quandt}}, \bibnamefont{and}
  \bibinfo{author}{\bibfnamefont{H.}~\bibnamefont{Weigel}},
  \emph{\bibinfo{title}{{Spectral Methods in Quantum Field Theory}}}, vol.
  \bibinfo{volume}{777} (\bibinfo{publisher}{Springer, New York},
  \bibinfo{year}{2009}).

\bibitem[{\citenamefont{{Alidad Askari, Aliakbar Moradi Marjaneh, Zhanna G.
  Rakhmatullina, Mahdy Ebrahimi-Loushab, Danial Saadatmand, Vakhid A. Gani,
  Panayotis G. Kevrekidis, and Sergey V. Dmitriev}}(2020)}]{ASKARI2020109854}
\bibinfo{author}{\bibnamefont{{Alidad Askari, Aliakbar Moradi Marjaneh, Zhanna
  G. Rakhmatullina, Mahdy Ebrahimi-Loushab, Danial Saadatmand, Vakhid A. Gani,
  Panayotis G. Kevrekidis, and Sergey V. Dmitriev}}}, \bibinfo{journal}{Chaos,
  Solitons and Fractals} \textbf{\bibinfo{volume}{138}},
  \bibinfo{pages}{109854} (\bibinfo{year}{2020}).

\bibitem[{\citenamefont{Saadatmand et~al.}(2018)\citenamefont{Saadatmand,
  Xiong, Kuzkin, Krivtsov, Savin, and Dmitriev}}]{PhysRevE.97.022217}
\bibinfo{author}{\bibfnamefont{D.}~\bibnamefont{Saadatmand}},
  \bibinfo{author}{\bibfnamefont{D.}~\bibnamefont{Xiong}},
  \bibinfo{author}{\bibfnamefont{V.~A.} \bibnamefont{Kuzkin}},
  \bibinfo{author}{\bibfnamefont{A.~M.} \bibnamefont{Krivtsov}},
  \bibinfo{author}{\bibfnamefont{A.~V.} \bibnamefont{Savin}}, \bibnamefont{and}
  \bibinfo{author}{\bibfnamefont{S.~V.} \bibnamefont{Dmitriev}},
  \bibinfo{journal}{Phys. Rev. E} \textbf{\bibinfo{volume}{97}},
  \bibinfo{pages}{022217} (\bibinfo{year}{2018}).

\bibitem[{\citenamefont{Evazzade et~al.}(2018)\citenamefont{Evazzade,
  Roknabadi, Behdani, Moosavi, Xiong, Zhou, and Dmitriev}}]{Evazzade2018-fi}
\bibinfo{author}{\bibfnamefont{I.}~\bibnamefont{Evazzade}},
  \bibinfo{author}{\bibfnamefont{M.~R.} \bibnamefont{Roknabadi}},
  \bibinfo{author}{\bibfnamefont{M.}~\bibnamefont{Behdani}},
  \bibinfo{author}{\bibfnamefont{F.}~\bibnamefont{Moosavi}},
  \bibinfo{author}{\bibfnamefont{D.}~\bibnamefont{Xiong}},
  \bibinfo{author}{\bibfnamefont{K.}~\bibnamefont{Zhou}}, \bibnamefont{and}
  \bibinfo{author}{\bibfnamefont{S.~V.} \bibnamefont{Dmitriev}},
  \bibinfo{journal}{Eur. Phys. J. B} \textbf{\bibinfo{volume}{91}}
  (\bibinfo{year}{2018}).

\bibitem[{\citenamefont{Dom\'{\i}nguez-Adame
  et~al.}(1995)\citenamefont{Dom\'{\i}nguez-Adame, S\'anchez, and
  Kivshar}}]{PhysRevE.52.R2183}
\bibinfo{author}{\bibfnamefont{F.}~\bibnamefont{Dom\'{\i}nguez-Adame}},
  \bibinfo{author}{\bibfnamefont{A.}~\bibnamefont{S\'anchez}},
  \bibnamefont{and} \bibinfo{author}{\bibfnamefont{Y.~S.}
  \bibnamefont{Kivshar}}, \bibinfo{journal}{Phys. Rev. E}
  \textbf{\bibinfo{volume}{52}}, \bibinfo{pages}{R2183} (\bibinfo{year}{1995}).

\bibitem[{\citenamefont{Speight}(1997)}]{Speight1997-hc}
\bibinfo{author}{\bibfnamefont{J.~M.} \bibnamefont{Speight}},
  \bibinfo{journal}{Nonlinearity} \textbf{\bibinfo{volume}{10}},
  \bibinfo{pages}{1615} (\bibinfo{year}{1997}).

\bibitem[{\citenamefont{Jaworski}(1987)}]{Jaworski.PLA.1987}
\bibinfo{author}{\bibfnamefont{M.}~\bibnamefont{Jaworski}}, \bibinfo{journal}{\
  Phys.\ Lett.\ A} \textbf{\bibinfo{volume}{125}}, \bibinfo{pages}{115}
  (\bibinfo{year}{1987}).

\bibitem[{\citenamefont{{F.~Hadipour, D.~Saadatmand, M.~Ashhadi,
  A.~Moradi~Marjaneh, I.~Evazzade, A.~Askari, and
  S.~V.~Dmitriev}}(2020)}]{Hadipour.PLA.2020}
\bibinfo{author}{\bibnamefont{{F.~Hadipour, D.~Saadatmand, M.~Ashhadi,
  A.~Moradi~Marjaneh, I.~Evazzade, A.~Askari, and S.~V.~Dmitriev}}},
  \bibinfo{journal}{Phys. Lett. A} \textbf{\bibinfo{volume}{384}},
  \bibinfo{pages}{126100} (\bibinfo{year}{2020}).

\bibitem[{\citenamefont{Weigel}(2017{\natexlab{a}})}]{Weigel:2016zbs}
\bibinfo{author}{\bibfnamefont{H.}~\bibnamefont{Weigel}},
  \bibinfo{journal}{Phys. Lett. B} \textbf{\bibinfo{volume}{766}},
  \bibinfo{pages}{65} (\bibinfo{year}{2017}{\natexlab{a}}).

\bibitem[{\citenamefont{Weigel}(2017{\natexlab{b}})}]{Weigel:2017iup}
\bibinfo{author}{\bibfnamefont{H.}~\bibnamefont{Weigel}},
  \bibinfo{journal}{Adv. High Energy Phys.} \textbf{\bibinfo{volume}{2017}},
  \bibinfo{pages}{1486912} (\bibinfo{year}{2017}{\natexlab{b}}).

\bibitem[{\citenamefont{Takyi and Weigel}(2016)}]{Takyi:2016tnc}
\bibinfo{author}{\bibfnamefont{I.}~\bibnamefont{Takyi}} \bibnamefont{and}
  \bibinfo{author}{\bibfnamefont{H.}~\bibnamefont{Weigel}},
  \bibinfo{journal}{Phys. Rev. D} \textbf{\bibinfo{volume}{94}},
  \bibinfo{pages}{085008} (\bibinfo{year}{2016}).

\bibitem[{\citenamefont{Xiong et~al.}(2017)\citenamefont{Xiong, Saadatmand, and
  Dmitriev}}]{PhysRevE.96.042109}
\bibinfo{author}{\bibfnamefont{D.}~\bibnamefont{Xiong}},
  \bibinfo{author}{\bibfnamefont{D.}~\bibnamefont{Saadatmand}},
  \bibnamefont{and} \bibinfo{author}{\bibfnamefont{S.~V.}
  \bibnamefont{Dmitriev}}, \bibinfo{journal}{Phys. Rev. E}
  \textbf{\bibinfo{volume}{96}}, \bibinfo{pages}{042109}
  (\bibinfo{year}{2017}).

\end{thebibliography}

\end{document}